\documentclass[sigconf]{acmart}
\DeclareMathOperator*{\argmax}{argmax}
\usepackage{hhline}
\usepackage{multirow}
\usepackage{threeparttable}
\usepackage{algorithm}
\usepackage{algorithmic}
\usepackage{array}
\usepackage{adjustbox}
\usepackage{caption}

\newcolumntype{L}[1]{>{\raggedright\let\newline\\\arraybackslash\hspace{0pt}}m{#1}}
\newcolumntype{C}[1]{>{\centering\let\newline\\\arraybackslash\hspace{0pt}}m{#1}}
\newcolumntype{R}[1]{>{\raggedleft\let\newline\\\arraybackslash\hspace{0pt}}m{#1}}

\usepackage{soul}

\usepackage{tikz}
\newcommand*\circled[1]{\tikz[baseline=(char.base)]{
            \node[shape=circle,draw,inner sep=0.5pt] (char) {#1};}}

\usepackage{colortbl}

\usepackage{hyperref}
\hypersetup{
    colorlinks=true,
    linkcolor=blue,
    filecolor=magenta,      
    urlcolor=cyan,
}

\AtBeginDocument{%
  \providecommand\BibTeX{{%
    \normalfont B\kern-0.5em{\scshape i\kern-0.25em b}\kern-0.8em\TeX}}}

\copyrightyear{2024}
\acmYear{2024}
\setcopyright{acmlicensed}\acmConference[SIGIR '24]{Proceedings of the 47th International ACM SIGIR Conference on Research and Development in Information Retrieval}{July 14--18, 2024}{Washington, DC, USA}
\acmBooktitle{Proceedings of the 47th International ACM SIGIR Conference on Research and Development in Information Retrieval (SIGIR '24), July 14--18, 2024, Washington, DC, USA}
\acmISBN{979-8-4007-0431-4/24/07}
\acmDOI{10.1145/3626772.3657710}

\begin{document}


\title[SyNCRec]{Pacer and Runner: Cooperative Learning Framework between Single- and Cross-Domain Sequential Recommendation}

\author{Chung Park}
\email{cpark88kr@gmail.com}
\affiliation{%
  \institution{SK Telecom, KAIST}
  \city{Seoul}
  \country{Republic of Korea}
  \country{ }
}

\author{Taesan Kim}
\email{ktmountain@sk.com}
\affiliation{%
  \institution{SK Telecom}
  \city{Seoul}
  \country{Republic of Korea}
  \country{ }
  }

\author{Hyungjun Yoon}
\email{hjyoon@sk.com}
\affiliation{%
  \institution{SK Telecom}
  \city{Seoul}
  \country{Republic of Korea}
  \country{ }
  }

\author{Junui Hong}
\email{skt.juhong@sk.com}
\affiliation{%
    \institution{SK Telecom, KAIST}
  \city{Seoul}
  \country{Republic of Korea}
  \country{ }
  }

\author{Yelim Yu}
\email{yelim.yu@sk.com}
\affiliation{%
  \institution{SK Telecom}
  \city{Seoul}
  \country{Republic of Korea}
  \country{ }
  }

\author{Mincheol Cho}
\email{skt.mccho@sk.com}
\affiliation{%
  \institution{SK Telecom}
  \city{Seoul}  
  \country{Republic of Korea}
  \country{ }
  }

\author{Minsung Choi}
\email{ms.choi@sk.com}
\affiliation{%
  \institution{SK Telecom}
  \city{Seoul}
  \country{Republic of Korea}
  \country{ }
  }

\author{Jaegul Choo}
\authornote{Corresponding Author (jchoo@kaist.ac.kr)}
\email{jchoo@kaist.ac.kr}
\affiliation{%
  \institution{Korea Advanced Institute of Science and Technology}
  \city{Daejeon}
  \country{Republic of Korea}
  \country{ }
  }

\renewcommand{\shortauthors}{Park Chung, et al.}


\begin{abstract}
Cross-Domain Sequential Recommendation (CDSR) improves recommendation performance by utilizing information from multiple domains, which contrasts with Single-Domain Sequential Recommendation (SDSR) that relies on a historical interaction within a specific domain.
However, CDSR may underperform compared to the SDSR approach in certain domains due to negative transfer, which occurs when there is a lack of relation between domains or different levels of data sparsity.
To address the issue of negative transfer, our proposed CDSR model estimates the degree of negative transfer of each domain and adaptively assigns it as a weight factor to the prediction loss, to control gradient flows through domains with significant negative transfer.
To this end, our model compares the performance of a model trained on multiple domains (CDSR) with a model trained solely on the specific domain (SDSR) to evaluate the negative transfer of each domain using our asymmetric cooperative network.
In addition, to facilitate the transfer of valuable cues between the SDSR and CDSR tasks, we developed an auxiliary loss that maximizes the mutual information between the representation pairs from both tasks on a per-domain basis.
This cooperative learning between SDSR and CDSR tasks is similar to the collaborative dynamics between pacers and runners in a marathon.
Our model outperformed numerous previous works in extensive experiments on two real-world industrial datasets across ten service domains.
We also have deployed our model in the recommendation system of our personal assistant app service, resulting in 21.4\% increase in click-through rate compared to existing models, which is valuable to real-world business\footnote{The implementation code of our model is available on {\color[HTML]{0037D7}\url{https://github.com/cpark88/SyNCRec}}.}.
\end{abstract}



\maketitle

\vspace{-0.1cm}
\section{Introduction}
In most real-world recommender systems, various business domains require understanding of the different interests and needs of users.
A practical approach to address this involves Single Domain Sequential Recommendation (SDSR), which focuses on recommending the next item within a specific domain using only the single-domain sequence and builds many domain-specific models  \cite{kang2018self,sun2019bert4rec, zhang2019feature, zhou2020s3}.
Another solution is Cross-Domain Sequential Recommendation (CDSR): it predicts the next item a user will interact with by leveraging their historical interaction sequences across multiple domains \cite{li2023one,cao2022contrastive,park2023cracking}.
In particular, CDSR has proven to be a promising approach to improve the performance of sequential recommendations across multiple domains simultaneously with one model \cite{li2023one,ma2022mixed,lin2023mixed, park2023cracking}. 
Note that the difference between CDSR and SDSR tasks for each domain lies in whether or not they use information from other domains in addition to the specific domain.

    \begin{figure}[h!]
    \begin{center}
    \includegraphics[width=1\linewidth]{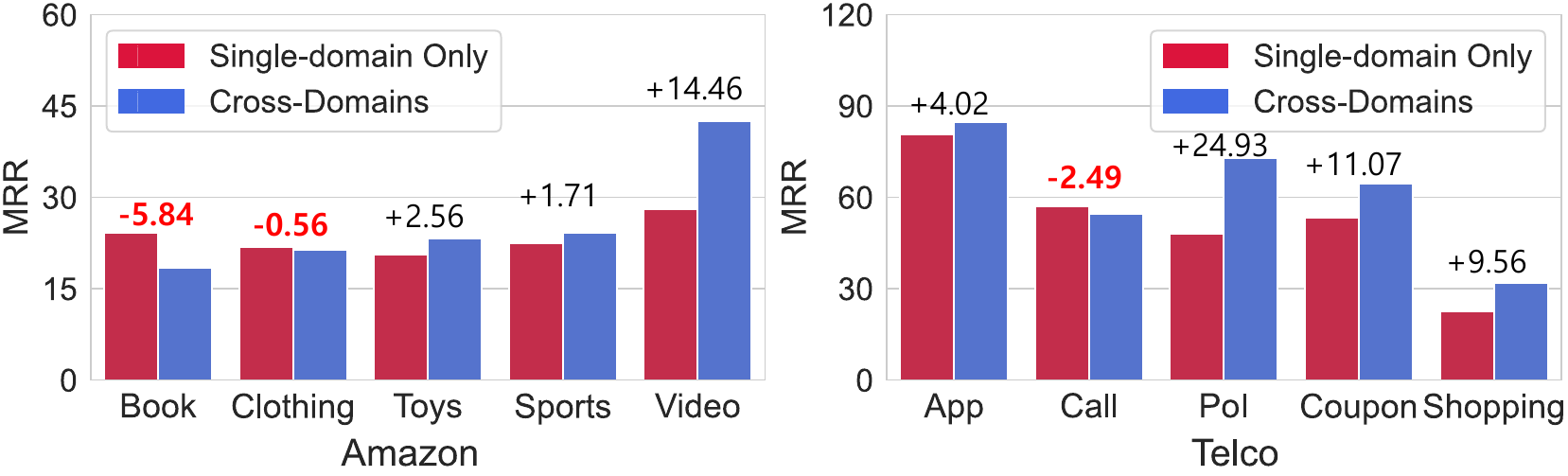}
    \end{center}
    \vspace{-0.3cm}
    \caption{In the \textit{Book}, \textit{Clothing} (Amazon dataset), and \textit{Call} (Telco dataset) domains, the SDSR (trained with a single domain) outperformed the CDSR approach (trained with multiple domains), which indicates negative transfer from other domains.
    We used CGRec \cite{park2023cracking} for the experiments. 
    }
    \label{fig:intro_2}
    \end{figure}

However, in certain domains, CDSR significantly underperforms relative to the SDSR approach, while in other domains it may outperform the SDSR approach.
This is because transferring knowledge from other weakly related domains or domains with different levels of data sparsity can have a negative impact on a particular domain: this is known as \textit{negative transfer} \cite{park2023cracking}.
To verify this experimentally, we measured the difference in the model performance between the CDSR and the SDSR approaches on a per-domain basis.
As illustrated in Fig.~\ref{fig:intro_2}, the recommendation performance of the $Book$ and $Clothing$ domains in the Amazon dataset, as well as the $Call$ domain in the Telco dataset, was worse when trained with other domains (i.e., CDSR) than when trained separately (i.e., SDSR).
This result is particularly reasonable for the $Book$ domain, considering the weak relationship between the $Book$ domain and other domains such as $Sports$ or $Toys$.
These experiments demonstrate that CDSR does not always outperform the SDSR approach across all domains.

In light of this, our primary goal is to improve the performance of the domains with significant negative transfer to ensure consistent high performance across multiple domains in the CDSR approach.
For this purpose, we propose the CDSR model \textbf{SyNCRec}, which stands for 
a A\underline{sy}mmetric Cooperative \underline{N}etwork for \underline{C}ross-Domain Sequential \underline{Rec}ommendation.
In SyNCRec, we assess the degree of negative transfer of each domain and adaptively assign this value as weight to the prediction loss corresponding to a specific domain.
This adaptive control of the gradient reduces its flow in domains with significant negative transfer, consequently enhancing the performance of domains adversely affected by other unrelated domains \cite{park2023cracking}.
To this end, our model estimates the negative transfer of a particular domain by comparing the performance of a model trained on domain-hybrid sequences (CDSR task) to the model trained solely on specific domain sequences (SDSR task) in our asymmetric cooperative network.
In the asymmetric cooperative network, the learning process of both tasks is partially decoupled to allow for proper calculation of their pure loss values without mutual interference, contributing to accurate computation of negative transfer.

In addition, to enhance the transfer of valuable clues across SDSR and CDSR tasks, we developed an auxiliary loss that maximizes the mutual information between the representation pairs from both tasks on a per-domain basis.
From this objective, we exploit the effective correlation signals inherent in the representation pairs of SDSR and CDSR tasks within a specific domain.
Our experimental results confirm that this auxiliary loss improves the positive transfer of information between the representations derived from the SDSR and CDSR tasks.

SDSR serves as a pacer to reduce negative transfer during the learning process of CDSR, similar to how pacers in marathons assist runners in avoiding a pace that is too fast or too slow.
As a result, SyNCRec improves the performance of the CDSR task compared to \textbf{25} state-of-the-art models, with two real-world datasets across ten domains.
Importantly, our model significantly improves the performance of domains that underperformed compared to the SDSR approach due to negative transfer.
In addition, extensive online experiments were conducted on large-scale recommender using industrial datasets to confirm its proficiency in real-world applications.
\ul{Note that our model can handle both SDSR and CDSR tasks simultaneously, allowing for the learning of multiple domains at once. This eliminates the need to build many domain-specific models, making it highly relevant for real-world scenarios.}

    \begin{figure*}[]
    \begin{center}
    \includegraphics[width=1\linewidth]{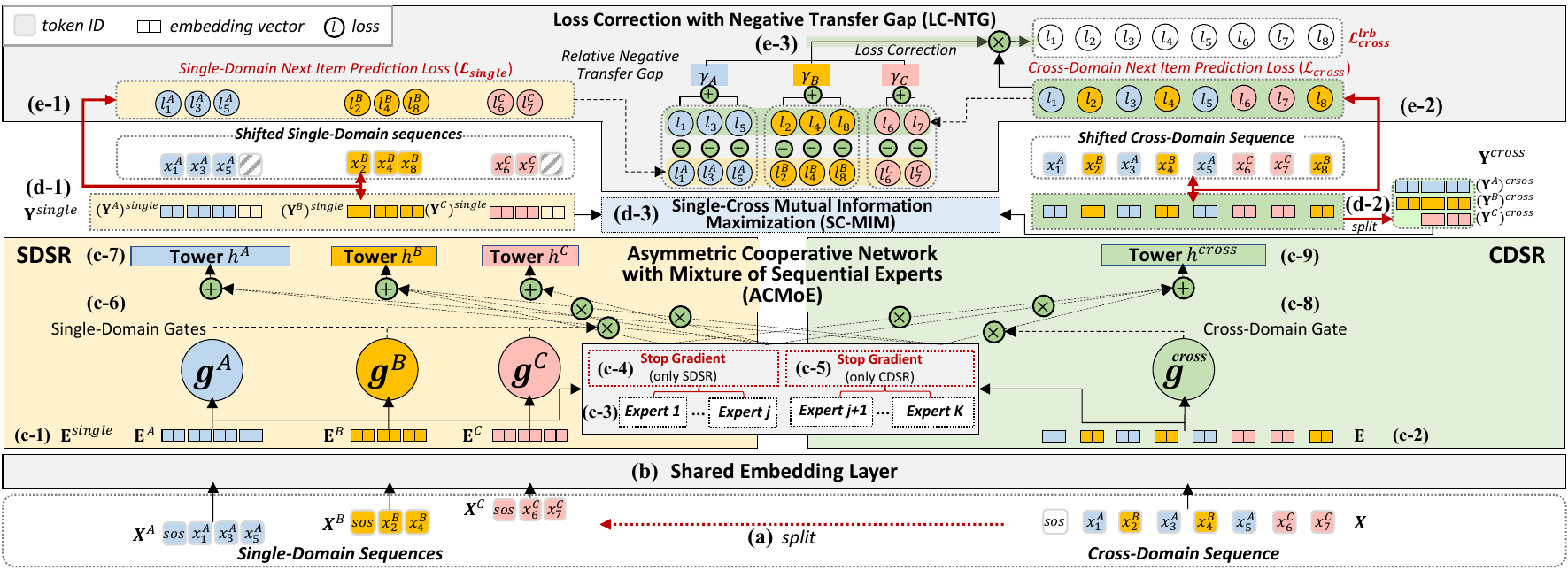}
    \end{center}
    \vspace{-0.3cm}
    \caption{
    We illustrate SyNCRec using three domains ($A$, $B$, $C$), each represented by a distinct color (blue, yellow, and pink).
    The notations $\oplus$, $\ominus$ and $\otimes$ indicate element-wise summation, subtraction, and multiplication, respectively.
    For the SDSR task for each domain, we input $X^{A}=[(\texttt{SOS}), x^{A}_{1},x^{A}_{3},x^{A}_{5}]$, $X^{B}=[(\texttt{SOS}), x^{B}_{2},x^{B}_{4}]$, and $X^{C}=[(\texttt{SOS}), x^{C}_{6}, x^{C}_{7}]$, and predict their shifted sequences $[x^{A}_{1},x^{A}_{3},x^{A}_{5},(\texttt{PAD})]$, $[x^{B}_{2},x^{B}_{4},x^{B}_{8}]$, and $[x^{C}_{6}, x^{C}_{7},(\texttt{PAD})]$. 
    For the CDSR task, we input $X=[(\texttt{SOS}), x^{A}_{1}, x^{B}_{2}, x^{A}_{3}, x^{B}_{4}, x^{A}_{5}, x^{C}_{6}, x^{C}_{7}]$ into the model, and predict the shifted sequence $[x^{A}_{1}, x^{B}_{2}, x^{A}_{3}, x^{B}_{4}, x^{A}_{5}, x^{C}_{6}, x^{C}_{7}, x^{B}_{8}]$.
    We do not consider losses for (\texttt{PAD}) tokens.
    }
    \label{fig:overall_model}
    \end{figure*}

\vspace{-0.16cm}
\section{Related Work} \label{section:related_work}
\subsection{Single-Domain Sequential Recommendation} \label{subsection:sr}
The SDSR framework is designed to model the temporal dynamics in user-item interactions, effectively capturing how user preferences evolve over time within a specific domain.
For example, GRU4Rec \cite{hidasi2015session}, STAMP \cite{liu2018stamp}, and NARM \cite{li2017neural} were introduced to predict the user's next item using GRU-based models.
Moreover, in SASRec \cite{kang2018self}, BERT4Rec \cite{sun2019bert4rec}, SINE \cite{tan2021sparse}, and LightSANs \cite{fan2021lighter}, attention mechanisms are utilized to encapsulate sequential patterns, thereby addressing both short- and long-term dependencies.
In addition, other machine learning architectures such as a convolutional network (e.g., NextItNet \cite{yuan2019simple}) or Markov Chain (e.g., TransRec \cite{he2017translation}) are adopted in the SDSR task.
Meanwhile, FDSA \cite{zhang2019feature}, S$^{3}$Rec \cite{zhou2020s3}, and CARCA \cite{rashed2022context} utilize item features (e.g., text description) in addition to item sequences to demonstrate their efficiency.
Nonetheless, studies on the SDSR framework continue to encounter issues of negative transfer in CDSR task.

\vspace{-0.25cm}
\subsection{Cross-Domain Sequential Recommendation}
The goal of Cross-Domain Recommendation (CDR) is to improve recommendations in a given domain by leveraging information from various other domains.
For example, CMF \cite{singh2008relational} and CLFM \cite{gao2013cross} modeled the matrix factorization of user-item interactions across multiple domains, capturing their shared patterns.
Other approaches, DTCDR \cite{zhu2019dtcdr}, DeepAPF \cite{yan2019deepapf}, and BiTGCF \cite{liu2020cross}, modeled user-item relationships in paired domains by a multi-task learning method.
However, these methods modeled pairwise relationships between domains for the CDR task, a strategy that may be less effective in scenarios involving a large number of domains.
To fill this research gap, CAT-ART \cite{li2023one} simultaneously modeled user and item representations in five domains using a matrix factorization approach. 
However, these studies have not taken into account the sequential nature of user interaction sequence.

Meanwhile, the CDSR aims to enhance the task of sequential recommendation, focusing on items that span multiple domains.
$\pi$-Net \cite{ma2019pi} introduced gating mechanisms designed to transfer information from a single domain to another paired domain. Meanwhile, C$^{2}$DSR \cite{cao2022contrastive} employed a self-attention based encoder and graph neural network to model both single- and cross-domain representations.
MIFN \cite{ma2022mixed} introduced the concept of mixed information flow, which reflects the knowledge flows between multiple domains.
MAN \cite{lin2023mixed} designed group-prototype attention mechanisms to capture domain-specific and cross-domain relationships.
However, these studies only modeled relationships between pairs of domains. In scenarios involving more than three domains, this approach would require handling an impractically large number of domain pairs, potentially limiting its feasibility in real-world applications \cite{li2023one}.
To address this limitation, CGRec \cite{park2023cracking} proposed a CDSR model that deals with more than three domains at once.
They approximated the negative transfer for each domain using Shapley values, and applied them to penalize items in domains with high negative transfer.
However, even in CGRec, we encountered problems with negative transfer, where some domains performed poorly over the SDSR approach (Fig. \ref{fig:intro_2}).
Additionally, CGRec showed the exponential computational complexity resulting from the calculation of Shapley values, which increases with the number of domains involved.
\ul{Therefore, our focus is on reducing negative transfer of all domains in the CDSR task while considering computational efficiency for rapid industrial deployment} (Section \ref{subsection: discusstion_negative_transfer} and \ref{subsection: discussion_model_variants}).

\vspace{-0.35cm}
\section{Preliminary}  \label{section:preliminary}
We consider an expanding CDSR task characterized by interaction sequences spanning more than \textbf{three} domains.
These domains may encompass diverse service platforms such as e-commerce or video platforms \cite{zhu2021cross,li2023one}.
Formally, the set of domains is denoted as $\mathcal{D}=\{A,B,C,...\}$, where $|\mathcal{D}|\ge3$.
We also consider the notation $d\in\mathcal{D}$ as a specific domain $d$.
The notation $V^{d}$ is introduced to represent a set of items specific to the domain $d$.
Then, $V$ indicates the total item set across all domains.

\noindent\textbf{Definition 1. Single- and Cross-Domain Sequential Recommendation}: The single-domain sequences of domain $d$ are represented by $X^{d} = [(\texttt{SOS}),$
$x_{1}^{d},x_{2}^{d}, ..., x_{|X^d|-1}^{d}]$, where each element $x_{t}^{d}$ signifies an interaction occurring at time $t$ and $(\texttt{SOS})$ is a start token of the sequence. 
We then define $X=(X^{A}, X^{B}, X^{C}, ...)$ as the cross-domain sequence where the $|\mathcal{D}|$ single-domain sequences are merged and rearranged in chronological order.  
Conversely, the cross-domain sequence $X$ can be split into $|\mathcal{D}|$ of $X^{d}$ (Fig. \ref{fig:overall_model}a). 
For example, consider a scenario where a user engages with three domains, labeled $\{A,B,C\}$. 
The cross-domain sequence for these domains is $X=[(\texttt{SOS}), x^{A}_{1}, x^{B}_{2}, x^{A}_{3}, x^{B}_{4},$
$ x^{A}_{5}, x^{C}_{6}, x^{C}_{7}]$, and is then split into $X^{A}=[(\texttt{SOS}), x^{A}_{1}$
$,x^{A}_{3},x^{A}_{5}]$, $X^{B}=[(\texttt{SOS}), x^{B}_{2},x^{B}_{4}]$, and $X^{C}=[(\texttt{SOS})$
$, x^{C}_{6}, x^{C}_{7}]$ from the split process.

\ul{Note that the SDSR solely focuses on recommending the next item within a specific domain $d$ using the single-domain sequence of domain $d$ (i.e., $X^{d}$) while the CDSR is to recommend the next item of domain $d$ based on the cross-domain sequence (i.e., $X$).}

\noindent\textbf{Definition 2. Negative Transfer Gap}: Based on previous studies \cite{zhang2023collaborative, wang2019characterizing}, we quantitatively measured the degree of negative transfer of each domain in the context of the CDSR task, referred to as the Negative Transfer Gap (NTG).
Suppose that $\mathcal{L}^d_{\pi}$ is the empirical loss of the sequential recommendation task using a model $\pi$ for a specific domain $d$, which takes the user's single-domain sequence $X^{d}$ (i.e., SDSR task) or the cross-domain sequence $X$ (i.e., CDSR task).
We can then define the negative transfer gap $\phi_{\pi}(d)$ of the domain $d$ as a quantifiable measure of negative transfer: $\phi_{\pi}(d)=\mathcal{L}^{d}_{\pi}(X^{d}) - \mathcal{L}^{d}_{\pi}(X)$, and \ul{negative transfer is identified when this value is negative.}
In Section \ref{subsection:loss_re_balancing}, we introduce the module that explicitly calculates the negative transfer gap $\phi_{\pi}(d)$ and we use it as the weighting factor for the domain-specific loss function.
 
\noindent\textbf{Problem Statement}: Given the historical cross-domain sequences $X_{1:t}$ observed up to the time step $t$, our goal is to predict the next item $x_{t+1}^{d}$:

\vspace{-0.5cm}
 \begin{equation}
  \begin{split}
 \label{equation:problem_statement}
  &\argmax_{x^{d}_{t+1}\in V^{d}} P\Big(x^{d}_{t+1} \Big\vert X_{1:t} \Big), 
 \end{split}
 \end{equation} 
where $P\Big(x^{d}_{t+1} \Big\vert X_{1:t} \Big)$ is the probability that the target item $x^{d}_{t+1}$ in domain $d$ will be interacted with by the given item sequences.
This goal is achieved through the SDSR and the CDSR tasks simultaneously.
In other words, there are $|\mathcal{D}|$ single-domain sequences for the SDSR and one cross-domain sequence for the CDSR task, and therefore $|\mathcal{D}|$+1 next item prediction tasks are performed in one model in a multi-task learning manner.

\section{Model} \label{section:model}
We illustrate the overall architecture of our model in Fig. \ref{fig:overall_model}.
\subsection{Shared Embedding Layer} \label{subsection:shared_embedding_layer}
During the embedding look-up phase, we acquire initialized representations of items for $|\mathcal{D}|$ single-domain sequences (i.e., $X^{d}$) and one cross-domain sequence (i.e., $X$).
For each domain $d$, we define an item embedding matrix ${M}^{d}$ represented by $\mathbb{R}^{|V^{d}| \times r}$, where $|V^{d}|$ indicates the total number of items in domain $d$, and $r$ is the embedding dimension.
We then concatenate the embedding matrices of all $|\mathcal{D}|$ domains to form an aggregated embedding matrix ${M}\in \mathbb{R}^{|V| \times r}$, where $|V|$ is the total number of item sets across all domains.

The single- and cross-domain sequences (i.e., $X^{d}$ and $X$) are subsequently converted into a fixed length, $T$.
If the sequence length is shorter than $T$, special tokens (\texttt{PAD}) are added to the left side, serving as placeholders for past interactions on $X^{d}$ and $X$.
Conversely, if the sequence length exceeds $T$, only the $T$ most recent actions are preserved.
As shown in Fig. \ref{fig:overall_model}b, the final embedding matrix $M$ is shared by the single- and cross-domain sequence by a look-up operation to extract the initialized sequences representation  ${\textbf{E}^{d}}\in \mathbb{R}^{T \times r}$ and ${\textbf{E}}\in \mathbb{R}^{T \times r}$, respectively (Fig. \ref{fig:overall_model}(c-1) and \ref{fig:overall_model}(c-2)).
The aggregation $|\mathcal{D}|$ of $\textbf{E}^d$ in chronological order can be described as $\textbf{E}^{single}$ (Fig. \ref{fig:overall_model}(c-1)).
Additionally, a learnable position embedding matrix $\textbf{P}\in \mathbb{R}^{T \times r}$ is integrated into $\textbf{E}^{d}$ and $\textbf{E}$ to encapsulate the sequential nature of user interaction ($\textbf{E}^{d} \leftarrow \textbf{E}^{d}+\textbf{P}$, $\textbf{E} \leftarrow \textbf{E}+\textbf{P}$) \cite{vaswani2017attention}.
The embedding representations of $\textbf{E}^{d}$ and $\textbf{E}$ at the $t$-th step are denoted as $e^{d}_{t}$ and $e_{t}$, respectively.

\vspace{-0.2cm}
\subsection{Asymmetric Cooperative Network with Mixture-of-Sequential Experts (ACMoE)}
Referring to previous studies \cite{zhang2023collaborative, wang2019characterizing}, the degree of the negative transfer of a specific domain can be defined as the difference between the $\circled{1}$ loss of the SDSR and $\circled{2}$ the loss of CDSR tasks for the domain (i.e., $ \circled{1}$ - $ \circled{2}$).
\ul{This value is called the \textbf{negative transfer gap (NTG)}, and helps determine whether samples from other domains are beneficial or detrimental to the specific domain.}
The degree of negative transfer in a specific domain increases as the NTG decreases, and therefore, we assign this value as the weight for the prediction loss in the domain, resulting in a smaller gradient flow.
To efficiently calculate this, we designed an asymmetric cooperative network that concurrently executes SDSR and CDSR tasks for each domain in a multi-task learning manner. 
For our multi-task learning, we utilized the multi-gate Mixture of Sequential Experts (MoE) architecture \cite{qin2020multitask}. 
This architecture explicitly models the relationships between different tasks and learns task-specific functionalities, enabling it to effectively leverage shared representations.
During this phase, we employed a decoupled mechanism allowing expert networks to be specialized in either SDSR or CDSR tasks, ensuring both tasks are implemented without mutual interference.
For this, in the decoupled mechanism, a stop-gradient operation is adopted 
on some experts for the SDSR tasks and on other experts for the CDSR task. 
This approach enables the losses for both tasks to be computed independently, leading to a more precise evaluation of the NTG.
We used Transformer \cite{vaswani2017attention} as the expert to handle sequential data.

\vspace{-0.2cm}
\subsubsection{Architecture} \label{subsubsection:mose}
Mathematically, given the initialized representations of single- and cross-domain sequences $\textbf{E}^{d}$ and $\textbf{E}$ from the shared embedding layer, we perform many-to-many sequence learning in each sequential expert.
In the case of single-domain sequences, for a given $T$-length inputs $\textbf{E}^{d}$, we can formulate the output of domain $d$ for this module as given below:

\vspace{-0.4cm}
 \begin{equation}
  \begin{split}
 \label{equation:mose_single}
  &(\textbf{Y}^{d})^{single} = h^{d}\Big(f^{d}(\textbf{E}^{d})\Big), \\ 
  &f^{d}(\textbf{E}^{d}) = \sum_{k=1}^{j} g^{d}(\textbf{E}^{d})_{k}  \texttt{SG}(f^{k}_{\texttt{TRM}}(\textbf{E}^{d})) + \sum_{k=j+1}^{K} g^{d}(\textbf{E}^{d})_{k}  f^{k}_{\texttt{TRM}}(\textbf{E}^{d})
 \end{split}
 \end{equation} 
where $h^{d}$ is the tower network for domain $d$ (Fig. \ref{fig:overall_model}(c-7)), $f^d$ is the multi-gated mixture of the sequential experts layer, $\texttt{SG}(\cdot)$ is the stop-gradient operation (Fig. \ref{fig:overall_model}(c-4)), and $f^{k}_{\texttt{TRM}}$ is the $k$-th transformer-based \cite{vaswani2017attention} sequential expert (Fig. \ref{fig:overall_model}(c-3); there are $K$ experts in total).
The tower network $h^{d}$ consists of a feed-forward network with layer normalization \cite{ba2016layer}.
The stop-gradient operation $\texttt{SG}(\cdot)$ serves as an identity function during the forward pass, but it drops the gradient for variables enclosed within it during the backward pass.
This operation is only used in the output from the $1$ to the $j$-th expert network ($j<K$).
Therefore, these experts cannot be updated by the single-domain sequences, but experts from $j$+$1$ to $K$ can learn the unique sequential pattern of single-domain sequences.
In other words, we can say that $K$-$j$ represents the number of experts trained by the SDSR task with the gradient descent algorithm.
We set $j$ to be 0.2 times the total number of experts $K$, and report a sensitivity analysis on the hyper-parameter $j$ in Section \ref{subsection:hyperparameter}.
In addition, $g^d$ is the gating network for domain $d$ (Fig. \ref{fig:overall_model}(c-6)), which converts the input into a distribution across the $K$ experts:
 \begin{equation}
  \begin{split}
 \label{equation:gating_single}
  &g^{d}(\textbf{E}^d) = softmax(W_{g}^{d}\textbf{E}^d)
 \end{split}
 \end{equation} 
where $W_{g}^{d}\in \mathbb{R}^{K \times dT}$ is the trainable fully-connected layer.
The aggregation $|\mathcal{D}|$ of $(\textbf{Y}^{d})^{single}$ in chronological order is described as $\textbf{Y}^{single}$ (Fig. \ref{fig:overall_model}(d-1)).
The $t$-th element of $\textbf{Y}^{single}$ is then $(y_{t}^{d})^{single}$.

We also perform the CDSR task using a similar scheme to that above for the single-domain case.
For a given $T$-length input $\textbf{E}$, we represent the output of ACMoE module as follows:
 \begin{equation}
  \begin{split}
 \label{equation:mose_cross}
  &\textbf{Y}^{cross} = h^{cross}\Big(f^{cross}(\textbf{E})\Big), \\ 
  &f^{cross}(\textbf{E}) = \sum_{k=1}^{j} g^{cross}(\textbf{E})_{k}  f^{k}_{\texttt{TRM}}(\textbf{E}) + \sum_{k=j+1}^{K} g^{cross}(\textbf{E})_{k}  \texttt{SG}(f^{k}_{\texttt{TRM}}(\textbf{E}))
 \end{split}
 \end{equation} 
where $h^{cross}$ is the tower network (Fig. \ref{fig:overall_model}(c-9)) and $f^{cross}$ is the multi-gated mixture of sequential experts layer for a cross-domain sequence.
The stop-gradient operation $\texttt{SG}(\cdot)$ (Fig. \ref{fig:overall_model}(c-5)) is only used in the output from $j$+1 to $K$-th $f^{k}_{\texttt{TRM}}$. 
From this operation, certain experts (from 1 to $j$) are able to learn the distinct sequential patterns present in cross-domain sequences.
In other words, we can say that $j$ is the number of experts trained by the CDSR task.
The gating network $g^{cross}(\textbf{E})=softmax(W_{g}^{cross}\textbf{E})$ for the cross-domain sequence maps the input to a distribution over the $K$ experts (Fig. \ref{fig:overall_model}(c-8)).
We can then formulate the output for this module  as $\textbf{Y}^{cross}= (y_1, y_2,...,y_T)^{cross}$.
\ul{Note that $(y_t^{d})^{single}$ and $(y_t)^{cross}$ are two representations of different views for the same item (i.e., also same domain $d$) at time step $t$, depending on whether it was encoded based on the single- or cross-domain sequence.}

\subsubsection{Transformer Experts} \label{subsubsection:transformer_experts}
We adopted the self-attention module introduced in Transformer \cite{vaswani2017attention} as the sequential expert layer $f^{k}_{\texttt{TRM}}$ (Fig. \ref{fig:overall_model}(c-3)), which consists of two distinct sub-layers as follows:
\noindent\textbf{Multi-head Self-Attention (MSA)}
At each MSA head, the inputs $Z \in \mathbb{R}^{T \times r}$ undergo a linear transformation resulting in three hidden representations, i.e., queries $Q_{i}\in \mathbb{R}^{T \times r/p}$, keys $K_{i}\in \mathbb{R}^{T \times r/p}$, and values $V_{i}\in \mathbb{R}^{T \times r/p}$, where $i$ indicates a specific head, $p$ is the number of heads, and $r$ is the dimension of the input.
Using these three hidden representations, the scaled dot-product attention ($\texttt{Attn}$) is then computed as follows:

\vspace{-0.4cm}
 \begin{equation}
 \begin{split}
 \label{equation:multi_head}
  &\texttt{Attn}(Q_{i}, K_{i}, V_{i}) = softmax(\frac{Q_{i}K_{i}^{\texttt{T}}}{\sqrt{r/p}})V_{i}, \\
  & Q_{i}=Z\mathrm{W}^{Q}_{i}, K_{i}=Z\mathrm{W}^{K}_{i}, V_{i}=Z\mathrm{W}^{V}_{i},
 \end{split}
 \end{equation}
where $\mathrm{W}^{Q}_{i}$,$\mathrm{W}^{K}_{i}$,$\mathrm{W}^{V}_{i} \in \mathbb{R}^{r \times r/p}$ are the trainable parameters. 
In the MSA, the above $\texttt{Attn}$ operation is implemented $p$ times in parallel, and then involves concatenating the outputs of each head and linearly projecting them to extract the final output $\textbf{H}\in \mathbb{R}^{T \times r}$:

\vspace{-0.4cm}
 \begin{equation}
 \begin{split}
 \label{equation:multi_head}
  &\textbf{H} = MSA(Z) = [\texttt{Attn}(Q_{1}, K_{1}, V_{1}) \vert\vert...\vert\vert \texttt{Attn}(Q_{p}, K_{p}, V_{p})]W^{F}
 \end{split}
 \end{equation}
where $\vert\vert$ is the concatenate operation and $W^F \in \mathbb{R}^{r \times r}$ is the trainable parameters. 
For the sequential recommendation, only information from prior time steps can be utilized.
Therefore, a masking operation is applied to the output of the MSA layer, effectively eliminating all connections between $Q_i$ and $K_j$ whenever $j>i$.

\noindent\textbf{Point-wise Feed-Forward Network (FFN)}
The FFN is then adopted into \textbf{H} to introduce nonlinearity and enable interactions among various latent subspaces using a fully-connected layer (FC), as follows:

\vspace{-0.2cm}
 \begin{equation}
 \begin{split}
 \label{equation:ffn}
  &\texttt{FFN}(\textbf{H}) = [\texttt{FC}(\textbf{H}_1) \vert\vert \texttt{FC}(\textbf{H}_2) \vert\vert,..., \vert\vert \texttt{FC}(\textbf{H}_T)], \\
  &\texttt{FC}(\textbf{H}_t)=GELU(\textbf{H}_{t}\mathrm{W}_{1}+b_{1})\mathrm{W}_{2}+b_{2},
 \end{split}
 \end{equation}
where $\textbf{H}_t$ is the $t$-th representation of $\textbf{H}$, $\mathrm{W}_1 \in \mathbb{R}^{r \times r}$, $b_1  \in \mathbb{R}^{r \times 1}$,$\mathrm{W}_2 \in \mathbb{R}^{r \times r}$, and $b_2 \in \mathbb{R}^{r \times 1}$ are learnable parameters, and $GELU$ is the gelu activation \cite{hendrycks2016gaussian}.

Therefore, the transformer-based sequential expert $f_{\texttt{TRM}}^{k}$ in Eq \ref{equation:mose_single} and \ref{equation:mose_cross} is described as follows: $f_{\texttt{TRM}}^{k}(\cdot) = \texttt{FFN}^{k}(\texttt{MSA}^{k}(\cdot))$.

\subsection{Loss Correction with Negative Transfer Gap (LC-NTG)} \label{subsection:loss_re_balancing}
From our ACMoE layer, we can explicitly measure the negative transfer gap (NTG) of each domain as described in Section \ref{section:preliminary}, and use the NTG to adaptively assign lower weights to the loss of items in domains that exhibit significant negative transfer during the training stage (Fig. \ref{fig:overall_model}(e-3)).

\subsubsection{Single-Domain Item Prediction}
As presented in Fig. \ref{fig:overall_model}(e-1), given the single-domain sequence $X^d_{1:t}$ and its expected next item $x_{t+1}^{d}$, we use the pairwise ranking loss \cite{rendle2012bpr} to optimize our model as follows: 

\vspace{-0.4cm}
 \begin{equation}
 \begin{split}
 \label{equation:single_objective}
  &l^{d}_{t} = log\sigma \Big(P(x_{t+1}^{d}=x^{d+} | X^{d}_{1:t}) - P(x_{t+1}^{d}=x^{d-} | X^{d}_{1:t})\Big),  \mathcal{L}^{d}_{single} = \sum_{t=1}^{T} l^{d}_{t},
 \end{split}
 \end{equation}
where $x^{d+}$ is the ground-truth item paired with a negative item $x^{d-}$ sampled from a uniform distribution, and $\sigma$ indicates the sigmoid activation function. 
Note that $P(x_{t+1}^{d}=x^{d} | X^{d}_{1:t}) $ is calculated as $ \sigma \Big( (y_{t}^{d})^{single} \cdot M(x^{d}) \Big)$, where $\cdot$ is the dot-product operation.

\subsubsection{Cross-Domain Item Prediction}
Similar to the single-domain item prediction, the CDSR task using the cross-domain sequence $X_{1:t}$ is performed with the following objective (Fig. \ref{fig:overall_model}(e-2)):

\vspace{-0.4cm}
 \begin{equation}
 \begin{split}
 \label{equation:cross_objective}
  &l_{t} = log\sigma \Big(P(x_{t+1}^{d}=x^{d+} | X_{1:t} ) - P(x_{t+1}^{d}=x^{d-} | X_{1:t})\Big),\; \mathcal{L}_{cross} = \sum_{t=1}^{T} l_{t},
 \end{split}
 \end{equation}
where $P(x_{t+1}^{d}=x^{d} | X_{1:t})$ is obtained by $\sigma \Big( (y_{t})^{cross} \cdot M(x^{d}) \Big)$.

\subsubsection{Calculating the Negative Transfer Gap}
We can describe the negative transfer gap (NTG) using the losses of Eq. \ref{equation:single_objective} and Eq. \ref{equation:cross_objective} as follows:

\vspace{-0.3cm}
 \begin{equation}
 \begin{split}
 \label{equation:negative_transfer_gap}
  & \phi_{\pi}(d) = \sum_{t=1}^{T}  \Big( l^{d}_{t} - l_{t}  \Big),
 \end{split}
 \end{equation}
where $l_t^d$ and $l_t$ are losses of the SDSR and CDSR tasks in time step $t$ for the domain $d$, respectively, calculated with our model $\pi$.
Note that in the cross-domain item prediction loss $l_t$, only the loss values corresponding to domain $d$ are used in Eq. \ref{equation:negative_transfer_gap}.

We then calculated the trainable relative NTG in each domain.
Let $\lambda_{d}$ represent the relative NTG of domain $d$, and $\lambda=(\lambda_{1},\lambda_{2},...,\lambda_{|\mathcal{D}|})$ be the vector of the relative NTG.
The vector of the relative NTG is initialized to be $\textbf{0}$.
From Eq. \ref{equation:negative_transfer_gap}, we obtain $\phi_{\pi}(d)$ for all domains at each training batch.
$\lambda$ is then updated in every batch as follows:
 \begin{equation}
 \label{equation:negative_transfer_vector}
 \lambda_{d} \leftarrow \texttt{softmax}(\alpha * \lambda_{d} +\beta * \phi_{\pi}(d); \; \delta); \;\; \forall d \in \mathcal{D},
 \end{equation}
where $\texttt{softmax}$$(;\delta)$ is the softmax function with the temperature $\delta$ \cite{szegedy2016rethinking}, and $\alpha$ and $\beta$ are learnable parameters.

\subsubsection{Loss Correction}
The relative NTG serves as a weight for the cross-domain item prediction loss computed in Eq. \ref{equation:cross_objective}.
This loss is re-aggregated by multiplying the relative NTG for each domain separately as follows (Fig. \ref{fig:overall_model}(e-3)):

\vspace{-0.4cm}
 \begin{equation}
 \label{equation:loss_rebalancing}
 \begin{split}
  &\mathcal{L}_{cross}^{lc} \\
  &= \sum_{t=1}^{T}\sum_{d=1}^{|\mathcal{D}|} \lambda_{d} \mathrm{log}\sigma\Bigg(P(x_{t+1}^{d}=x^{d+}|X_{1:t})  -P(x_{t+1}^{d}=x^{d-}|X_{1:t})\Bigg).
 \end{split}
 \end{equation}
\ul{This adaptive regulation of the gradient diminishes its flow in domains where there is significant negative transfer, thereby alleviating negative transfer effects across all domains.}

\subsection{Single-Cross Mutual Information Maximization (SC-MIM)} 
To improve the transfer of valuable information between SDSR and CDSR tasks, we developed the Single-Cross Mutual Information Maximization (SC-MIM), an auxiliary loss that maximizes the mutual information between the representation pairs from the SDSR and CDSR tasks on a per-domain basis.
The SC-MIM extracts the correlation signals between the intrinsic characteristics of the both tasks for a specific domain.
This module is an application of the concept of mutual information, which reflects how knowledge of one random variable reduces uncertainty in the other random variable.
Mathematically, the mutual information $I$ between random variables $X$ and $Y$ is given as follows:

\vspace{-0.4cm}
 \begin{equation}
 \begin{split}
 \label{equation:mim}
  & I(X, Y) = D_{KL}(p(X,Y) \vert\vert p(X)p(Y)) = \mathbb{E}_{p(X,Y)} \Bigg[ log \frac{p(X,Y)}{p(X)p(Y)} \Bigg],
 \end{split}
 \end{equation}
where $D_{KL}(\cdot \vert\vert \cdot)$ is the Kullback–Leibler divergence. 
However, maximizing mutual information in high-dimensional spaces can be a challenging task. Therefore, in practice, it is common to maximize a tractable lower bound on $I(X, Y)$.
A specific lower bound that has demonstrated effective practical results is InfoNCE \cite{kong2019mutual, zhou2020s3}, which is defined as follows:

\vspace{-0.4cm}
 \begin{equation}
 \begin{split}
 \label{equation:info_nce}
  &I(X,Y) \geq \mathbb{E}_{p(X,Y)}[\rho_{\theta}(x,y) - \mathbb{E}_{q(\hat{Y})}(log\sum_{\hat{y}\in\hat{Y}} \text{exp} \; \rho_{\theta}(x, \hat{y}))] + log\vert\hat{Y}\vert,
 \end{split}
 \end{equation}
where $x$ and $y$ are two distinct viewpoints of the same input, and $\rho_{\theta}$ represents a parameterized similarity function with $\theta$.
And $\hat{Y}$ is a set of samples chosen from a proposal distribution $q(\hat{Y})$, which includes a positive sample $y$ and $\vert\hat{Y}\vert$-1 negative samples.
Refer to the detailed derivations of InfoNCE (Eq. \ref{equation:info_nce}) as a lower bound for mutual information in \citet{oord2018representation}.
If $\hat{Y}$ contains all possible outcomes of the uniformly distributed random variable $Y$, then maximizing InfoNCE is equal to maximizing the standard cross-entropy loss:

\vspace{-0.4cm}
 \begin{equation}
 \begin{split}
 \label{equation:info_nce_uniform}
  &\mathbb{E}_{p(X,Y)} [ \rho_{\theta}(x,y) - log \sum_{\hat{y}\in Y} \text{exp} \; \rho_{\theta} (x,\hat{y})].
 \end{split}
 \end{equation}

In our problem, we focus on maximizing the mutual information of single- and cross-domain representations (i.e., $\textbf{Y}^{single}$ and $\textbf{Y}^{cross}$), which are different views of the user sequence.
For this, we split the cross-domain representation $\textbf{Y}^{cross}$ into a subsequence per domain $(\textbf{Y}^{d})^{cross}$ (Fig. \ref{fig:overall_model}(d-2)).
Then, as shown in Fig. \ref{fig:overall_model}(d-3), we can describe the mutual information between single- and cross-domain sequences for domain $d$ as follows: 

\vspace{-0.2cm}
 \begin{equation}
 \begin{split}
 \label{equation:mim_ours}
  &\mathcal{L}_{SC-MIM}^{d} = \rho(\;(\textbf{Y}^{d})^{single}, (\textbf{Y}^{d})^{cross}\;) \\ 
  &\;\;\;\;\;\;\;\;\;\;\;\; -log \sum_{u-} \text{exp}(\rho(\;(\textbf{Y}^{d})^{single-}, (\textbf{Y}^{d})^{cross})\;) ,
 \end{split}
 \end{equation}
where $u-$ is the other users in a training batch, and $(\textbf{Y}^{d})^{single-}$ is the subsequence of domain $d$ of user $u-$.
In addition, $\rho(\cdot, \cdot)$ is implemented according to the following equation:

\vspace{-0.4cm}
 \begin{equation}
 \begin{split}
 \label{equation:mim_details}
  &\rho(U, V) = \sigma (U^{\texttt{T}} \cdot W^{H} \cdot V),
 \end{split}
 \end{equation}
where $\sigma$ is the sigmoid function and $W^{H} \in \mathbb{R}^{r \times r}$ is a trainable parameter matrix. 

\vspace{-0.1cm}
\subsection{Model Training and Evaluation} \label{subsection: training_obj}
The total training loss function of our model is as follows:

\vspace{-0.3cm}
 \begin{equation}
 \label{equation:total_training}
 \begin{split}
  &\mathcal{L} = \eta \Bigg( \sum_{d=1}^{|\mathcal{D}|} \Big(\mathcal{L}^{d}_{single} \Big) + \mathcal{L}_{cross}^{lc} \Bigg) + (1-\eta) \sum_{d=1}^{|\mathcal{D}|}\mathcal{L}_{SC-MIM}^{d},
 \end{split}
 \end{equation}
where $\eta$ is the harmonic factor.
\ul{In the evaluation stage, we only used cross-domain representations to make predictions.}
For example, given the latest representations $(y_{T})^{cross}$, the next recommended item is selected based on the highest prediction score in domain $d$:

\vspace{-0.3cm}
 \begin{equation}
  \begin{split}
 \label{equation:evaluation}
  &\argmax_{x^d\in V^{d}} \big( (y_{T})^{cross})^{\texttt{T}} \cdot M(x^d) \big),
 \end{split}
 \end{equation} 
where $V^d$ is the item set of domain $d$ and $M$ is the embedding layer.

\section{Experiments}
The experiments are designed to answer the following research questions:

\noindent\textbf{(RQ1)}: Does the performance of our model surpass the current state-of-the-art baselines in practical applications that involve more than three domains? 

\noindent\textbf{(RQ2)}: Can our model effectively address the challenge of negative transfer across all domains in the CDSR task?

\noindent\textbf{(RQ3)}: What is the impact of various components of our model on its performance in CDSR tasks? 

\noindent\textbf{(RQ4)}: How do variations in hyper-parameter settings influence the performance of our model?

\noindent\textbf{(RQ5)}: How does the model perform when deployed online?

\begin{table}[]
\caption{Statistics of datasets}\label{tab:data_summary}
\vspace{-0.3cm}
    \renewcommand{\arraystretch}{0.85}
\begin{adjustbox}{width=\columnwidth,center}
\begin{tabular}{c|ccccc}
\hline\hline
\textbf{Dataset}                 & \textbf{\#Users}                  & \textbf{Domain}     & \textbf{\#Items} & \textbf{\#Interactions} & \textbf{Sparsity} \\ \hline
\multirow{5}{*}{\textbf{Amazon}} & \multirow{5}{*}{105,364} & Books      & 425,985 & 1,422,676      & 99.98\%  \\
                        &                          & Clothing   & 290,804 & 947,417        & 99.98\%  \\
                        &                          & Video      & 29,013  & 292,891        & 99.97\%  \\
                        &                          & Toys       & 121,559 & 575,449        & 99.98\%  \\
                        &                          & Sports     & 133,066 & 541,717        & 99.99\%  \\ \hline
\multirow{5}{*}{\textbf{Telco}}  & \multirow{5}{*}{99,936}  & App-Use    & 9,192   & 36,716,069     & 96.01\%  \\
                        &                          & Call-Use   & 3,301   & 3,038,385      & 99.08\%  \\
                        &                          & Navi       & 549     & 1,892,868      & 96.44\%  \\
                        &                          & Coupon-Use & 72      & 401,065        & 92.08\%  \\
                        &                          & e-comm     & 714     & 230,233        & 99.60\%  \\ \hline\hline

\end{tabular}
\end{adjustbox}
\end{table}

\vspace{-0.2cm}
\subsection{Datasets}
\subsubsection{\textbf{Amazon Dataset}} 
Amazon review datasets \cite{mcauley2015image} was used in our experiments, encompassing five domains: \textit{Books}, \textit{Clothing Shoes and Jewelry}, \textit{Video and Games}, \textit{Toys and Games}, and \textit{Sports and Outdoors} (Table \ref{tab:data_summary}).  
These domains are abbreviated as \textit{Books}, \textit{Clothing}, \textit{Video}, \textit{Toys}, and \textit{Sports}, respectively.

\vspace{-0.1cm}
\subsubsection{\textbf{Industrial Dataset (Telco)}} 
We collected user logs from diverse real-world applications operated by a leading global telecommunications company. 
The dataset comprised customers who consented to the collection and analysis of their data. 
The data contain five domains: Application Usage (\textit{APP-Use}), Call Record (\textit{Call-Use}), Navigation Service (\textit{Navi}), e-commerce service (\textit{e-comm}), and Discount Coupon Usage (\textit{Coupon-Use}).
\ul{We identified 99,936 users, and there are 48,887 non-overlapping users across five domains.}

\begin{table*}[h] \footnotesize
\centering
\caption{Overall model performance comparison in Amazon dataset. 
The top two methods are highlighted in bold and underlined.
}
    \vspace{-0.35cm}
    \setlength{\tabcolsep}{1.33pt}
    \renewcommand{\arraystretch}{0.88}
\label{tab:result_1}
\begin{tabular}{c|ccccc|ccccc|ccccc|ccccc|ccccc}
\hline\hline
\textbf{Domain}                  & \multicolumn{5}{c|}{\textbf{Books}}                       & \multicolumn{5}{c|}{\textbf{Clothing}}                    & \multicolumn{5}{c|}{\textbf{Toys}}                       & \multicolumn{5}{c|}{\textbf{Sports}}                       & \multicolumn{5}{c}{\textbf{Video}}                       \\ \hline\hline
\multirow{2}{*}{\textbf{Models}} & \multicolumn{2}{c}{HR} & \multicolumn{2}{c}{NDCG} & MRR   & \multicolumn{2}{c}{HR} & \multicolumn{2}{c}{NDCG} & MRR   & \multicolumn{2}{c}{HR} & \multicolumn{2}{c}{NDCG} & MRR   & \multicolumn{2}{c}{HR} & \multicolumn{2}{c}{NDCG} & MRR  & \multicolumn{2}{c}{HR} & \multicolumn{2}{c}{NDCG} & MRR   \\ \cline{2-26} 
                                 & @5        & @10        & @5         & @10         & @10   & @5        & @10        & @5          & @10        & @10   & @5        & @10        & @5         & @10         & @10   & @5        & @10        & @5         & @10         & @10  & @5        & @10        & @5         & @10         & @10   \\ \hline
BPRMF                            &    0.262       &    0.361        &    0.191        &   0.223         &  0.180     &      0.225     &    0.337        &  0.152 &  0.188  &  0.143           & 0.286           &   0.403    &    0.201       &  0.238    & 0.188     &    0.274        &       0.390      &  0.189     &    0.226       &   0.176        &   0.537         &      0.658       &    0.421  &      0.460     &   0.398              \\
GCMC                             &     0.309      &      0.412      &      0.227      &      0.260       &   0.214    &    0.318       &   0.429         &    0.221         &     0.257       &  0.204     &   0.382        &  0.500          &    0.274        &     0.312        &   0.254    &     0.377      &   0.491         &    0.269        &     0.306        &  0.248    &    \underline{0.659}       &   0.750         &  \underline{0.535}          &    0.564         &   0.505    \\ \hline
GRU4Rec                          &    0.265       &     0.348       &     0.195       &    0.222         &  0.183     &   0.334        &    0.434        &   0.242          &    0.274        &   0.224    &  0.344         &   0.451         &     0.242       &     0.277        &   0.223    &     0.356      &   0.464         &     0.256       &     0.291        &   0.238   &    0.551       &     0.661       &    0.416        &   0.452          &    0.386   \\
SASRec                           &   0.261        &     0.355       &      0.187      &     0.218        &  0.193     &    0.306       &     0.415       &   0.219         &    0.255        &  0.223     &     0.345      &     0.473       &     0.243       &     0.284        &    0.243   &    0.352       &     0.476       &    0.251        &    0.291         &    0.250  &   0.623        &    0.744        &   0.474         &     0.514        &  0.452     \\
BERT4Rec                         &   0.208        &    0.296        &    0.144        &   0.172          &  0.134     &     0.277      &   0.387         &      0.192       &    0.228        &   0.179    &    0.311       &     0.433       &    0.214        &     0.254        &  0.199     &    0.320       &    0.442       &    0.225        &   0.264          &  0.210    &     0.549      &    0.676        & 0.415           &   0.456          &   0.387    \\
S3Rec                            &   0.236        &    0.316        &     0.173       &    0.199         & 0.164      &    0.336       &    0.442        &   \underline{0.248}          &      \underline{0.282}      &  0.233     &  0.335         &      0.442      &     0.242       &    0.276         &    0.225   &    0.382       &   0.490         &    0.283        &    0.318         &   0.265    &    0.500       &     0.612       &  0.386          &    0.423         &    0.364   \\
CARCA                            &    0.240       &    0.303        &      0.183      &    0.203         &  0.190     &     0.229      &    0.303        &     0.168        &     0.192       &   0.180    &   0.278        &    0.406        &   0.206         &      0.192       &    0.214   &    0.277       &    0.390        &    0.202        &     0.238        &  0.214    &   0.559        &  0.666          &   0.438        &      0.473       &   0.425    \\
NextItNet                        &     0.222      &   0.303         & 0.155           &    0.181         &  0.143     &   0.285        &      0.383      &    0.200         &   0.232         &   0.185    &    0.325       &     0.432       &    0.227        &      0.262       &   0.209    &   0.326       &   0.433         &    0.231        &      0.265       &  0.213    &      0.535     &  0.655          &    0.395        &      0.434       &   0.364    \\
SINE                             &   0.282        &  0.380          &  0.198          &       0.230      &  0.184     & 0.297          &  0.411       &0.208 &0.244       & 0.194            & 0.383            & 0.500      & 0.279          & \underline{0.317}           &  0.260          &  0.376           &  0.487     & 0.273          & 0.309           & 0.254           &    \underline{0.682}         &0.772      &  0.550         &0.579            & 0.518               \\
STAMP                            &     0.256      & 0.340           &  0.185          &               0.212  &  0.173         & 0.333           & 0.433            & 0.242           & 0.274      & 0.225          & 0.353           &  0.465          &  0.251           &  0.287     & 0.232          &  0.374          & 0.478           &  0.274           &  0.307    &    0.254       & 0.587           & 0.693           & 0.456            &  0.491 & 0.427     \\
TransRec                         &  0.261         & 0.343           &  0.192          & 0.219            & 0.181      & 0.288          &  0.389          &  0.207           &  0.239          &  0.193     &  0.349         &  0.456          &  0.250          &  0.285           &  0.232     &  0.341         & 0.444           &  0.247          &  0.281           &  0.230    &  0.578         &  0.667          & 0.465           &  0.494           &  0.439     \\
LightSANs                        &    0.264       &  0.353          &    0.193        &  0.222           & 0.181      & 0.262          & 0.365           & 0.189            & 0.222           &  0.179     &  0.346         & 0.461           & 0.247           & 0.284            &  0.230     &  0.331         & 0.443           & 0.239           &  0.275           &  0.223    &   0.605        &  0.710          & 0.477           &  0.511           &  0.449     \\
NARM                             &      0.254     &  0.340          &  0.182          &  0.210           & 0.169      &  0.307         &  0.413          & 0.218            &  0.252          & 0.203      &  0.337         &  0.455          &  0.237          &  0.275           & 0.220      &  0.348         &  0.463          & 0.249           &  0.286           & 0.232     &  0.540         & 0.661           & 0.413           &  0.452           &  0.387     \\ \hline
BiTGCF                           &    0.289       & 0.376           & 0.215           &0.243             & 0.202      &  0.200         &  0.277          &  0.144           &  0.169          & 0.136      & 0.219          &  0.309          &  0.155          &  0.184           & 0.146      &  0.200         &  0.281          & 0.144           &  0.170           & 0.136     &  0.317         & 0.437           & 0.231           & 0.269            &  0.218     \\
DTCDR                            & 0.288          &  0.368          &  0.212          & 0.244            & 0.206      &  0.262         &   0.360         &  0.181           &   0.213         &  0.168     & 0.268          & 0.373           &  0.188          &  0.222           &  0.176     &  0.285         & 0.391           &  0.200          &  0.234           &  0.186    &  0.391         &  0.536          & 0.273           & 0.320            &  0.254     \\
CMF                              &  0.288         & 0.365           & 0.221           &   0.246          &   0.209    & 0.247          &  0.333          &  0.179           & 0.206           & 0.168      &  0.234         &  0.334          & 0.164           &  0.196           & 0.154      &  0.228         & 0.311           & 0.163           & 0.190            & 0.153     & 0.304          & 0.399           & 0.226           & 0.256            & 0.212      \\
CLFM                             &     0.244      & 0.320           & 0.182           & 0.206            & 0.172      & 0.200          &  0.276          & 0.140            &  0.165          & 0.131      &  0.218         &  0.313          &  0.153          &  0.184           & 0.145      & 0.195          & 0.268           & 0.140           & 0.163            & 0.131     &  0.285         &   0.395         &  0.200          & 0.236            &  0.187     \\
CAT-ART                          &     0.314      &  0.411          & 0.250           &  0.270           & \textbf{0.258}      &  0.267         &  0.348          &  0.212           & 0.242           & 0.203      & 0.265          &  0.355          & 0.209           & 0.261            &  0.226     &  0.236         &  0.355          &  0.152          &  0.245           &  0.160    &  0.359         & 0.470           & 0.244           & 0.308            & 0.259      \\
DeepAFP                          &  0.271         & 0.350           & 0.207           & 0.232            & 0.196      & 0.224          & 0.308           & 0.159            & 0.186           & 0.149      & 0.155          & 0.215           & 0.110           & 0.129            & 0.103      & 0.154          & 0.209           & 0.113           & 0.130            & 0.107     & 0.196          & 0.261           & 0.145           & 0.166            &  0.137     \\ \hline
MIFN                             &    0.219       & 0.293           & 0.160           & 0.184            & 0.150      & 0.229          & 0.323           &  0.164           & 0.194           & 0.155      & 0.235          & 0.333           & 0.165           & 0.196            & 0.155      & 0.253          & 0.342           & 0.182           & 0.211            & 0.170     & 0.398          & 0.531           &  0.279          & 0.322            &  0.258     \\
$\pi$-net                           &   0.236        & 0.319           &  0.176          &  0.176           & 0.167      &  0.220         &  0.317          & 0.154            & 0.185           & 0.145      & 0.236          & 0.331           & 0.164           & 0.195            & 0.153      & 0.245          & 0.337           & 0.172           & 0.201            & 0.160     &  0.390         & 0.523           & 0.270           & 0.314            & 0.249      \\
MMoE                             &     0.236      & 0.344           & 0.170           & 0.205            & 0.182      & 0.288          & 0.405           & 0.202            & 0.240           & 0.207      & 0.276          & 0.419           & 0.189           &  0.235           & 0.199      &  0.280         & 0.419           &0.195            & 0.239            &  0.205    &  0.351         &  0.599          & 0.235           & 0.315            &  0.251     \\
MAN                              &   0.315        &  0.374          & 0.244           & \underline{0.274}            & 0.240      &  0.261         &  0.304          &  0.138           &  0.227          & 0.184      &  0.267         &  0.309          &  0.170          & 0.259            &  0.203     &  0.238         & 0.342           &  0.138          & 0.245            & 0.156     & 0.363          & 0.462           & 0.246           & 0.312            &  0.270     \\
C$^{2}$DSR                            &      \underline{0.320}     &  \underline{0.420}          & \underline{0.241}           &  0.271           & 0.243      & 0.234          & 0.345           & 0.182            & 0.218           & 0.174      &  0.265         &  0.379          &  0.184          &  0.221           & 0.194      &  0.256         & 0.367           &  0.182          & 0.218            & 0.193     &  0.362         & 0.501           & 0.249           &  0.294           &  0.250     \\ 
CGRec          &  0.261         & 0.352           & 0.189           & 0.218            & 0.195      & \underline{0.339}          & \underline{0.452}           & 0.242            & 0.278           &  \underline{0.242}     & \underline{0.384}           & \underline{0.504}           & \underline{0.274}           & 0.313            & \underline{0.270}      &  \underline{0.394}         & \underline{0.512}           &  \underline{0.284}          & \underline{0.323}            & \underline{0.280}     &  0.655         & \underline{0.758}           & 0.519           & 0.553            & \underline{0.497}                      \\ \hline\hline
\textbf{Ours}                             &  \textbf{0.337}         & \textbf{0.433}           & \textbf{0.249}           & \textbf{0.280}             & \underline{0.248}       & \textbf{0.366}          & \textbf{0.480}            & \textbf{0.262}             & \textbf{0.298}           & \textbf{0.258}       & \textbf{0.442}          & \textbf{0.567}           & \textbf{0.320}           &  \textbf{0.361}           &  \textbf{0.312}     &  \textbf{0.438}         & \textbf{0.553}           & \textbf{0.317}           & \textbf{0.355}            & \textbf{0.307}     & \textbf{0.724}          & \textbf{0.810}           & \textbf{0.597}           & \textbf{0.625}            &  \textbf{0.574}     \\ \hline\hline
\end{tabular}
\end{table*}

\begin{table*}[h] \footnotesize
\centering
\caption{Overall model performance comparison in Telco dataset. 
The top two methods are highlighted in bold and underlined.
}
    \vspace{-0.35cm}
    \setlength{\tabcolsep}{1.33pt}
    \renewcommand{\arraystretch}{0.88}
\label{tab:result_2}
\begin{tabular}{c|ccccc|ccccc|ccccc|ccccc|ccccc}
\hline\hline
\textbf{Domain}                  & \multicolumn{5}{c|}{\textbf{App-Use}}                     & \multicolumn{5}{c|}{\textbf{Call-Use}}                    & \multicolumn{5}{c|}{\textbf{Navi}}                        & \multicolumn{5}{c|}{\textbf{Coupon-Use}}                   & \multicolumn{5}{c}{\textbf{e-comm}}                       \\ \hline\hline
\multirow{2}{*}{\textbf{Models}} & \multicolumn{2}{c}{HR} & \multicolumn{2}{c}{NDCG} & MRR   & \multicolumn{2}{c}{HR} & \multicolumn{2}{c}{NDCG} & MRR   & \multicolumn{2}{c}{HR} & \multicolumn{2}{c}{NDCG} & MRR  & \multicolumn{2}{c}{HR} & \multicolumn{2}{c}{NDCG} & MRR    & \multicolumn{2}{c}{HR} & \multicolumn{2}{c}{NDCG} & MRR   \\ \cline{2-26} 
                                 & @5        & @10        & @5         & @10         & @10   & @5        & @10        & @5          & @10        & @10   & @5        & @10        & @5          & @10        & @10  & @5        & @10        & @5         & @10         & @10    & @5        & @10        & @5         & @10         & @10   \\ \hline
BPRMF                            &  0.945         &  0.973          &  0.865          &   0.874          & 0.842      &    0.647       &    0.807        &   0.452          &  0.504          &  0.409     & 0.850          &    0.946        &    0.666         &    0.697        &   0.617   &  0.873         &       0.937     &    0.687        &      0.707       &  0.632      &      0.128     &  0.355          &  0.062          &  0.134           &  0.070     \\
GCMC                             & 0.939          & 0.972           & 0.845           & 0.856            & 0.817      & 0.628          & 0.806           &0.431             & 0.489           &  0.390     &0.836           & 0.938           & 0.642            & 0.675           & 0.592     &  0.870         & 0.935           & 0.664           & 0.686            & 0.604       & 0.214          & 0.422           & 0.114           & 0.181            &  0.109     \\ \hline
GRU4Rec                          &   0.946        & 0.976           & 0.842           &  0.852           & 0.811      &0.691           & 0.841           & 0.505            &  0.554          & 0.463      & 0.904          & 0.970           &     0.744        &  0.766          & 0.700     & 0.871          & 0.944           & 0.665           & 0.689            & 0.606       & 0.350          & 0.511           & 0.249           & 0.301            & 0.236      \\
SASRec                           & 0.952          & 0.980           &  0.858          & 0.867            & 0.829      & 0.713          & 0.852           & 0.544            & 0.589           & 0.506      & 0.905          & 0.972           & 0.753            & 0.775           &  0.711    & 0.879          & 0.949           &  0.687          &  0.710           & 0.633       & 0.391          & 0.527           & 0.290           & 0.334            &  0.274     \\
BERT4Rec                         & 0.945          & 0.976           & 0.839           & 0.849            & 0.807      & 0.651          & 0.816           & 0.448            & 0.501           & 0.403      & 0.872          & 0.955           & 0.678            & 0.705           & 0.624     & 0.849          &  0.942          &  0.616          & 0.646            &0.551        & 0.290          &  0.497          &  0.170          &  0.236           &  0.157     \\
S3Rec                            & 0.953          & 0.979           & 0.858           & 0.867            & 0.830      & 0.714          & 0.849           & 0.546            &  0.590          &  0.509     & \underline{0.925}          & 0.978           & 0.777            &  0.795          &  0.735    &    0.884       &  0.949          & 0.691           & 0.712            & 0.635       & 0.406          &  0.557          &  0.299          &  0.347           &  0.284     \\
CARCA                            &  0.815         &  0.836          &  0.747          & 0.754            &  0.430     &  0.315         & 0.406           &  0.213           & 0.243           &  0.205     &  0.739         &  0.834          &  0.545           & 0.575           & 0.498     &  0.860         & 0.912           & 0.662           &  0.680           & 0.606       & 0.053          & 0.131           &  0.027          &  0.052           &     0.050  \\
NextItNet                        &    0.937       &  0.972          &  0.826          & 0.837            & 0.793      & 0.669          & 0.828           &  0.471           & 0.523           &  0.427     &  0.882         &  0.962          &  0.709           & 0.735           & 0.661     &  0.849         & 0.933           & 0.636           & 0.663            & 0.576       & 0.312          & 0.472           &  0.214          &  0.265           &  0.202     \\
SINE                             & 0.926          & 0.964           & 0.811           & 0.824            & 0.777      &  0.581         & 0.759           & 0.381            & 0.438           & 0.339      &  0.829         &  0.928          &  0.612           & 0.644           & 0.553     &    0.844       & 0.912           &    0.618 & 0.640 & 0.551       &  0.099           &  0.285      & 0.045          & 0.104      & 0.052       \\
STAMP                            &  0.950         & 0.978           & 0.854           & 0.863            & 0.825      &0.690           & 0.839           & 0.508            & 0.557           &  0.468     & 0.898          & 0.968           & 0.735            & 0.758           & 0.690     &  0.879         &  0.945          &0.682            &  0.704           & 0.625       &  0.346         &  0.492          & 0.244           & 0.291            &  0.229     \\
TransRec                         & 0.931          &  0.969          &  0.821          & 0.834            & 0.789      & 0.636          & 0.799           & 0.466            & 0.519           & 0.432      & 0.874          & 0.957           &  0.698           &  0.725          & 0.650     & 0.857          & 0.918           &  0.654          & 0.675            & 0.595       & 0.320          & 0.428           & 0.237           &      0.271       &  0.223     \\
LightSANs                        & 0.957          &  \underline{0.980}          & 0.870           & 0.878            & 0.844      &  0.722         & 0.858           &  0.555           &  0.600          & 0.518      &  0.921         & \underline{0.976}           & 0.778            & 0.796           & 0.737     & 0.884          & 0.951           & 0.699           & 0.722            & 0.647       &  0.385         & 0.542           & 0.280           &  0.330           &  0.265     \\
NARM                             & 0.949          & 0.977           & 0.849           & 0.859            & 0.819      &  0.687         & 0.836           &  0.491           & 0.540           & 0.447      &  0.907         &  0.970          & 0.743            &  0.763          &  0.696    &  0.878         & 0.948           &  0.668          &  0.691           & 0.607       &  0.331         & 0.492           & 0.232           &  0.284           &  0.220     \\ \hline
BiTGCF                           &  0.953         &  0.977          & 0.861           & 0.869            & 0.834      &  0.613         & 0.755           & 0.455            & 0.501           & 0.422      & 0.607          & 0.749           &  0.459           & 0.505           & 0.429     &  0.860         &  0.920          &  0.758          & 0.778            &  0.732      & 0.376          & 0.495           & 0.292           & 0.331            & 0.281      \\
DTCDR                            &  0.941         &  0.973          &  0.838          & 0.849            & 0.808      &  0.629         & 0.771           & 0.474            & 0.521           & 0.442      & 0.606          &  0.759          &  0.467           &  0.516          & 0.441     &  0.815         & 0.913           & 0.611           & 0.643            &  0.556      &  0.250         & 0.430           & 0.155           & 0.213            &  0.148     \\
CMF                              & \textbf{0.962}          &  \textbf{0.981}          & 0.881           &  0.887           & 0.856      & 0.625          &  0.772          & 0.479            & 0.526           & 0.450      & 0.633          & 0.772           & 0.505            & 0.550           & 0.482     &  0.847         &  0.907          & 0.762           & 0.782            & 0.742       &  0.302         &  0.419          & 0.239           & 0.277            &  0.234     \\
CLFM                             & \underline{0.959}          &    0.979        & 0.873           & 0.880            & 0.847      & 0.654          & 0.792           & 0.495            &  0.540          & 0.462      &  0.594         &  0.741          & 0.468            & 0.516           & 0.446     &  0.829         &  0.888          &  0.741          &  0.760           & 0.720       & 0.277          & 0.380           & 0.219           &  0.253           &  0.214     \\
CAT-ART                          &     0.952      &  0.972          & 0.872           & 0.886            & 0.848      & 0.727          & 0.815           & 0.574            &  0.624          & \underline{0.569}      & 0.639          & 0.783           &  0.508           & 0.568           &  0.468    &      0.868     & 0.929           & 0.763           & 0.777            & 0.732       &  0.395         & 0.518           & 0.295           &  0.341           &  0.295     \\
DeepAFP                          & 0.959          &  0.977          &  \textbf{0.889}          &  \textbf{0.895}           & \textbf{0.867}      &  0.590         & 0.744           & 0.452            & 0.501           &  0.427     & 0.611          & 0.754           & 0.491            & 0.537           & 0.470     & 0.835          & 0.895           & 0.751           &0.771             & 0.732       & 0.277          & 0.372           &  0.225          & 0.255            &  0.220     \\ \hline
MIFN                             &  0.886         & 0.940           & 0.763           & 0.781            & 0.729      & 0.688          & 0.792           & 0.578            & 0.612           & 0.556      & 0.661          &  0.773          & 0.538            & 0.574           & 0.511     & 0.761          & 0.870           & 0.642           & 0.677            & 0.618       &  0.417         & 0.608           & 0.313           & 0.374            &  0.303     \\
$\pi$-net                           & 0.884          & 0.939           & 0.763           & 0.781            &  0.729     & 0.665          & 0.772           & 0.555            & 0.590           &  0.533     &  0.671         & 0.781           & 0.547            & 0.583           & 0.521     & 0.752          &  0.860          & 0.637           & 0.672            &  0.614      & 0.435          & 0.575           & 0.320           & 0.365            &  0.301     \\
MMoE                             & 0.887          &  0.927          & 0.817           & 0.830            & 0.803      &  0.374         & 0.405           & 0.343            &0.354            & 0.351      & 0.357          & 0.373           & 0.334            & 0.339           & 0.347     & 0.245          & 0.327           & 0.202           & 0.228            & 0.230       & 0.292          & 0.333           &  0.231          &  0.244           & 0.235      \\
MAN                              &  0.922         & 0.967           & 0.847           & 0.866            & 0.811      & 0.683          & 0.777           & 0.533            & 0.565           & 0.510      & 0.664          & 0.764           & 0.500            & 0.591           & 0.549     & 0.803          & 0.891           &0.667            & 0.705            & 0.661       & 0.360          & 0.528           & 0.280           & 0.337            & 0.296      \\

C$^{2}$DSR                            & 0.936          & 0.964           & 0.866           & 0.869            & 0.813      & 0.682          & 0.790           & 0.566            & 0.601           & 0.552      & 0.677          & 0.784           & 0.547            & 0.587           & 0.536     & 0.803          &0.895            &0.685            & 0.715            & 0.665       & \underline{0.420}          & \underline{0.571}           & 0.305           &   0.354          &  0.306     \\ 
CGRec                            & 0.951          &  0.977          &  0.871          & 0.880            & 0.849      & \underline{0.744}          & \underline{0.861}           & \underline{0.590}            & \underline{0.628}           &  0.562     &  0.920         & 0.972           & \underline{0.791}            & \underline{0.808}           & \underline{0.757}     &  \underline{0.912}         & \underline{0.965}           & 0.756           &0.774             & 0.713       & 0.417          & 0.568           & \underline{0.318}           & \underline{0.366}            & \underline{0.325}      \\ \hline\hline
\textbf{Ours}                             &    0.956       & 0.979           & \underline{0.885}           & \underline{0.893}            & \underline{0.866}      &  \textbf{0.833}         & \textbf{0.916}           & \textbf{0.689}            & \textbf{0.716}           & \textbf{0.651}      & \textbf{0.968}          & \textbf{0.992}           & \textbf{0.878}            & \textbf{0.886}            & \textbf{0.851}     & \textbf{0.971}          & \textbf{0.990}           & \textbf{0.895}           & \textbf{0.901}            & \textbf{0.872}       & \textbf{0.583}          & \textbf{0.751}           & \textbf{0.451}           & \textbf{0.505}            & \textbf{0.443}      \\ \hline\hline
\end{tabular}
\end{table*}

\vspace{-0.2cm}
\subsection{Experimental Settings}
\subsubsection{Baseline Models}
The performance of our model was evaluated with four categories of baseline models: (1) General Recommendation (\textbf{GR}), (2) Single-Domain Sequential Recommendation (\textbf{SR}), (3) Cross-Domain Recommendation (\textbf{CDR}), and (4) Cross-Domain Sequential Recommendation (\textbf{CDSR}), as shown in Tables \ref{tab:result_1} and \ref{tab:result_2}. 
The baselines in the GR include BPRMF \cite{rendle2012bpr} and GCMC \cite{berg2017graph}. 
BPRMF is a classical approach that develops a pairwise loss function for modeling the relative preferences of users.
GCMC proposes a graph autoencoder framework to address the matrix completion. 
Refer to Section \ref{section:related_work} for a description of SR, CDR, and CDSR baselines.
The RecBole framework \cite{zhao2021recbole,recbole[2.0]} was used to implement all CDR baselines and some SR models.
CARCA \cite{rashed2022context}, S$^{3}$Rec \cite{zhou2020s3}, C$^{2}$DSR \cite{cao2022contrastive}, MIFN \cite{ma2022mixed}, CAT-ART \cite{li2023one}, $\pi$-net \cite{ma2019pi}, and MAN \cite{lin2023mixed} were executed with their official code.
For the implementation of CGRec \cite{park2023cracking}, we made use of the source code that the authors provided.

\subsubsection{Evaluation Settings and Metrics}
In line with prior studies \cite{kang2018self, zhou2020s3, rashed2022context}, we employed the \textit{leave-one-out} approach for evaluating recommendation performance. 
Specifically, each user interaction sequence was divided into three parts: the last item designated as test data, the penultimate item reserved for validation, and all preceding items used as training data.
Performance for each domain was assessed independently, depending on the domain of the last item in the test data, resulting in domain-specific performance metrics derived from different user groups.

Given the vast number of items, using all items as test candidates was impractical due to time constraints. 
Consequently, a widely adopted method was implemented, where the ground truth item (positive sample) is paired with 99 randomly chosen items (negative samples) that the user had not engaged with before.
We then reported the performance of Top-$k$ recommendations, which is based on a ranked list of 100 items. 
The assessment focused on several key metrics: \textit{Hit Ratio} (HR), \textit{Normalized Discounted Cumulative Gain} (NDCG), and \textit{Mean Reciprocal Rank} (MRR).
Note that the performance of GR, SR, CDR, and CDSR baselines including ours was measured using cross-domain sequences on the CDSR scenario. 

\vspace{-0.3cm}
\subsubsection{Implementation Details} \label{subsubsection: hyperparameter_setting}
The size of the embedding ($r$), training batch, and maximum sequence length ($T$) were all set to 128. 
The number of experts in ACMoE is set to four for the Amazon dataset and five for the Telco dataset.
The transformer layer in the expert network utilizes two single-head attention blocks, each with a head size of four.
For updating all parameters, the Adam optimizer was employed.
To guarantee a fair comparison, we configured the baseline model's sequence length, embedding size, batch size, and optimizer to match those of our model.

\vspace{-0.2cm}
\subsection{Performance Evaluation (RQ1)}
We evaluated the performance of predicting the next item among all baselines and our model. 
The results for the two datasets are reported in Table \ref{tab:result_1} (\textbf{Amazon}) and Table \ref{tab:result_2} (\textbf{Telco}). 
Based on these results, we can draw the following conclusions.

\noindent \textbf{(1) The effectiveness of our model can be observed}.  
Our model outperforms most baseline models on two real-world datasets.\footnote[2]{All CDR and CDSR baselines, except CAT-ART and CGRec, were designed to model only pairwise relationships between domains. 
Therefore, for these baselines, two out of five domains were chosen for training. 
In other words, each domain was paired with the other four domains, and the average performance is reported in Tables \ref{tab:result_1} and \ref{tab:result_2}. Due to space constraints, we cannot provide a detailed pairwise performance analysis of these CDR and CDSR models across domains.} 
Our model improved upon the best baseline model (second best for each domain) by \textbf{+3.13\%} (\textit{Book}), \textbf{+1.18\%} (\textit{Clothing}), \textbf{+7.29\%} (\textit{Video}), \textbf{+9.65\%} (\textit{Toys}), and \textbf{+5.28\%} (\textit{Sports}) for HR@5 on the Amazon dataset, and \textbf{+11.96\%} (\textit{Call}-\textit{Use}), \textbf{+4.65\%} (\textit{Navi}), \textbf{+6.47\%} (\textit{Coupon}-\textit{Use}), and \textbf{+38.81\%} (\textit{e}-\textit{comm}) on the Telco dataset.
In the \textit{App}-\textit{Use} domain, our model did not achieve the top performance, but the difference from the best-performing baseline (i.e., CMF) was minimal, at only -0.63\%.
Our model thus performs consistently well across all domains, unlike some other models such as CMF and DeepAFP which show high performance in certain domains but significantly lower performance in others.


\noindent \textbf{(2) Integrating information from all domains simultaneously in a model can improve performance in each domain compared to modeling a pairwise domain-domain relationship.}
CDSR baselines that modeled five domains simultaneously, such as CAT-ART, CGRec, and our model, outperformed other baselines that only trained on domain pairs (e.g., C$^{2}$DSR, MAN, MIFN, and $\pi$-net).
Furthermore, extending these domain pair-wise CDSR models to a multi-domain CDSR scenario with $|\mathcal{D}|$ domains would require managing at least $|\mathcal{D}| \choose 2$ pairs of relations, which becomes impractical when the number of domains is substantial (See footnote 2).
As a result, we confirmed the effectiveness of combining information from various domains at once for the CDSR task.

\vspace{-0.25cm}
\subsection{Discussion of the negative transfer (RQ2)} \label{subsection: discusstion_negative_transfer}
The phenomenon of negative transfer becomes apparent when comparing the performance of  the domain-specific model, trained exclusively on single-domain sequence, with the performance of baseline models trained on cross-domain sequences.
For the domain-specific model, we used CGRec \cite{park2023cracking}, a state-of-the-art CDSR model that can also be trained on a single-domain sequence.
As shown in Table \ref{tab:result_3}, all models, except for ours, performed worse than the SDSR approach for two to nine of the ten domains in both datasets.
Our model outperformed the SDSR approach for all domains, particularly in the \textit{Clothing}, \textit{Toys}, and \textit{Call}-\textit{Use} domains, where other models performed poorly.
\textbf{In conclusion, our model significantly mitigates the negative transfer problem compared to other CDSR baselines for all domains.}

\begin{table}[] \footnotesize
\centering
\caption{The performance improvement(\%) of the representative CDSR baselines trained on the cross-domain sequences compared to the CGRec \cite{park2023cracking} trained on the single-domain sequence only. If this value is positive, then it is highlighted in {\color[HTML]{0037D7} blue}; otherwise {\color[HTML]{CB0000} red}.}
    \vspace{-0.35cm}
    \setlength{\tabcolsep}{0.8pt}
    \renewcommand{\arraystretch}{0.9}
\label{tab:result_3}
\begin{tabular}{c|c|ccccc|ccccc}
\hline\hline
\multirow{2}{*}{\textbf{Models}} & \textbf{Dataset} & \multicolumn{5}{c|}{\textbf{Amazon}}     & \multicolumn{5}{c}{\textbf{Telco}}             \\ \cline{2-12} 
                                 & Domains          & B & C & V & T & S & A & C & N & M & E \\ \hline
\multirow{3}{*}{DTCDR}            & HR@5             &  {\color[HTML]{CB0000} -13.7}     &  {\color[HTML]{CB0000} -20.9}        &   {\color[HTML]{CB0000} -14.0}    & {\color[HTML]{CB0000} -15.7}     & {\color[HTML]{CB0000} -8.6 }        & {\color[HTML]{CB0000} -0.6 }        & {\color[HTML]{CB0000} -14.9 }          & {\color[HTML]{CB0000} -10.0 }    & {\color[HTML]{CB0000} -0.5 }           & {\color[HTML]{CB0000} -32.3 }       \\
                                 & NDCG@5           & {\color[HTML]{CB0000} -16.8 }      & {\color[HTML]{CB0000} -23.3 }         & {\color[HTML]{CB0000} -14.4 }     & {\color[HTML]{CB0000} -17.3 }     & {\color[HTML]{CB0000} -9.6 }       & {\color[HTML]{CB0000} -0.2 }        & {\color[HTML]{CB0000} -21.3 }         & {\color[HTML]{CB0000} -9.4 }     &  {\color[HTML]{0037D7} +2.7}          & {\color[HTML]{CB0000} -36.4}       \\ 
          & MRR           & {\color[HTML]{CB0000} -14.9 }      & {\color[HTML]{CB0000} -23.5 }         & {\color[HTML]{CB0000} -14.7 }      & {\color[HTML]{CB0000} -17.6 }     & {\color[HTML]{CB0000} -9.4 }       & {\color[HTML]{0037D7} 0.0}        & {\color[HTML]{CB0000} -22.5 }         & {\color[HTML]{CB0000} -8.3}     &  {\color[HTML]{0037D7} +4.2}          & {\color[HTML]{CB0000} -34.3 }       \\ \hline
\multirow{3}{*}{CAT-ART}         & HR@5             & {\color[HTML]{CB0000} -6.0 }      & {\color[HTML]{CB0000} -19.3 }         & {\color[HTML]{CB0000} -14.9 }      & {\color[HTML]{CB0000} -30.2 }     & {\color[HTML]{CB0000} -16.2 }       &  {\color[HTML]{0037D7} +0.5}       & {\color[HTML]{CB0000} -1.6 }         & {\color[HTML]{CB0000} -5.2 }    & {\color[HTML]{0037D7} +6.0 }           & {\color[HTML]{0037D7} +7.0 }       \\
                                 & NDCG@5           &    {\color[HTML]{CB0000} -2.0 }   & {\color[HTML]{CB0000} -10.2 }         & {\color[HTML]{CB0000} -5.2 }       &  {\color[HTML]{CB0000} -37.2 }    & {\color[HTML]{CB0000} -19.3 }       & {\color[HTML]{0037D7} +3.8 }        & {\color[HTML]{CB0000} -4.8 }         & {\color[HTML]{CB0000} -1.5 }    &  {\color[HTML]{0037D7} +28.4 }          &  {\color[HTML]{0037D7} +21.2 }      \\ 
          & MRR           & {\color[HTML]{0037D7} +6.5 }      & {\color[HTML]{CB0000} -7.3 }         & {\color[HTML]{0037D7} +9.5 }      & {\color[HTML]{CB0000} -29.0 }     & {\color[HTML]{CB0000} -7.6 }       & {\color[HTML]{0037D7} +5.0  }      & {\color[HTML]{CB0000} -0.3 }         & {\color[HTML]{CB0000} -2.6 }     &  {\color[HTML]{0037D7} +37.1  }        & {\color[HTML]{0037D7} +31.3  }     \\ 
                                 \hline
\multirow{3}{*}{MAN}           & HR@5             &   {\color[HTML]{CB0000} -5.6}    & {\color[HTML]{CB0000} -21.3}         & {\color[HTML]{CB0000} -14.4 }      & {\color[HTML]{CB0000} -29.6 }     &  {\color[HTML]{CB0000} -15.2 }      & {\color[HTML]{CB0000} -2.7 }        & {\color[HTML]{CB0000} -7.6 }         & {\color[HTML]{CB0000} -1.4 }    & {\color[HTML]{CB0000} -2.0 }           & {\color[HTML]{CB0000} -2.4 }       \\
                                 & NDCG@5           & {\color[HTML]{CB0000} -4.2 }      & {\color[HTML]{CB0000} -41.5 }         & {\color[HTML]{CB0000} -22.6 }      &  {\color[HTML]{CB0000} -42.8}    & {\color[HTML]{CB0000} -18.5}       & {\color[HTML]{0037D7} +0.8  }      &  {\color[HTML]{CB0000} -11.6 }        &   {\color[HTML]{CB0000} -3.0 } &     {\color[HTML]{0037D7} +12.1   }    & {\color[HTML]{0037D7} +14.8 }       \\ 
          & MRR           & {\color[HTML]{CB0000} -0.7}      & {\color[HTML]{CB0000} -16.1 }         & {\color[HTML]{CB0000} -1.5 }      & {\color[HTML]{CB0000} -30.6 }     & {\color[HTML]{CB0000} -3.7 }       & {\color[HTML]{0037D7} +0.4 }        & {\color[HTML]{CB0000} -10.7 }         & {\color[HTML]{0037D7} +14.2 }     &  {\color[HTML]{0037D7} +23.8   }       & {\color[HTML]{0037D7} +31.7 }      \\                                  
                                 \hline
\multirow{3}{*}{C$^{2}$DSR}           & HR@5             &   {\color[HTML]{CB0000} -4.3 }    & {\color[HTML]{CB0000} -29.5 }         & {\color[HTML]{CB0000} -15.0 }      & {\color[HTML]{CB0000} -24.4 }     & {\color[HTML]{CB0000}  -15.5}       & {\color[HTML]{CB0000} -1.1 }        &      {\color[HTML]{CB0000} -7.6 }    & {\color[HTML]{0037D7} +0.4 }    &  {\color[HTML]{CB0000} -2.0 }          & {\color[HTML]{0037D7} +14.0   }    \\
                                 & NDCG@5           &    {\color[HTML]{CB0000} -5.7 }  & {\color[HTML]{CB0000} -22.8 }         & {\color[HTML]{CB0000} -16.1 }      & {\color[HTML]{CB0000} -24.6  }    & {\color[HTML]{CB0000} -17.6  }      &  {\color[HTML]{0037D7} +3.1  }     & {\color[HTML]{CB0000} -6.0  }        & {\color[HTML]{0037D7} +6.2 }    &  {\color[HTML]{0037D7} +15.3}          & {\color[HTML]{0037D7} +25.5  }     \\ 
          & MRR           & {\color[HTML]{0037D7} +0.2 }     & {\color[HTML]{CB0000} -20.8   }       & {\color[HTML]{CB0000} -6.2  }     &{\color[HTML]{CB0000} -14.3 }     & {\color[HTML]{CB0000} -10.7   }     & {\color[HTML]{0037D7}+0.7}        & {\color[HTML]{CB0000} -3.3 }         & {\color[HTML]{0037D7} +11.5 }    & {\color[HTML]{0037D7} +24.6}          & {\color[HTML]{0037D7} +36.1 }       \\                                  
                                 \hline
\multirow{3}{*}{CGRec}           & HR@5             & {\color[HTML]{CB0000} -24.8 }      &  {\color[HTML]{CB0000} -5.4 }        & {\color[HTML]{0037D7} +20.1}       & {\color[HTML]{0037D7} +10.7 }     & {\color[HTML]{0037D7} +54.2 }       & {\color[HTML]{0037D7} +0.4 }        &   {\color[HTML]{0037D7} +0.7 }       &  {\color[HTML]{0037D7} +36.5 }   &  {\color[HTML]{0037D7} +11.4 }          & {\color[HTML]{0037D7} +13.0 }       \\
                                 & NDCG@5           &   {\color[HTML]{CB0000} -29.3 }    & {\color[HTML]{CB0000} -5.0 }         & {\color[HTML]{0037D7} +21.9 }      & {\color[HTML]{0037D7} +12.4 }     & {\color[HTML]{0037D7} +71.8 }       & {\color[HTML]{0037D7} +3.7 }        & {\color[HTML]{CB0000} -2.1 }          & {\color[HTML]{0037D7} +53.5 }     & {\color[HTML]{0037D7} +27.2 }           &  {\color[HTML]{0037D7} +30.6 }      \\ 
          & MRR           & {\color[HTML]{CB0000} -23.1 }      & {\color[HTML]{CB0000} -3.2 }         & {\color[HTML]{0037D7} +28.8 }      & {\color[HTML]{0037D7} +19.6  }    & {\color[HTML]{0037D7} +77.0 }       & {\color[HTML]{0037D7} +5.1 }        & {\color[HTML]{CB0000} -1.5  }        & {\color[HTML]{0037D7} +57.5 }     &  {\color[HTML]{0037D7}+33.7}          & {\color[HTML]{0037D7} +44.8 }       \\                                  
                                 \hline\hline
\multirow{3}{*}{Ours}            & HR@5             &   {\color[HTML]{0037D7} +1.0 }    & {\color[HTML]{0037D7} +10.6 }         & {\color[HTML]{0037D7} +42.0  }     & {\color[HTML]{0037D7} +29.7  }    & {\color[HTML]{0037D7} +69.2  }      & {\color[HTML]{0037D7} +0.9 }        & {\color[HTML]{0037D7} +12.7 }         & {\color[HTML]{0037D7} +43.7  }   & {\color[HTML]{0037D7} +18.5  }          & {\color[HTML]{0037D7}+58.1  }       \\
                                 & NDCG@5           & {\color[HTML]{CB0000} -2.6 }      & {\color[HTML]{0037D7} +10.8  }         & {\color[HTML]{0037D7}  +45.7  }    &    {\color[HTML]{0037D7} +31.3 }  &  {\color[HTML]{0037D7} +97.6 }     & {\color[HTML]{0037D7} +5.4  }       & {\color[HTML]{0037D7} +14.3   }        & {\color[HTML]{0037D7} +70.4  }    &  {\color[HTML]{0037D7} +50.5  }         & {\color[HTML]{0037D7} +85.5  }      \\ 
          & MRR           & {\color[HTML]{0037D7} +2.3   }    & {\color[HTML]{0037D7} +17.6 }         & {\color[HTML]{0037D7} +50.8  }     & {\color[HTML]{0037D7} +36.4 }     & {\color[HTML]{0037D7} +104.9  }      & {\color[HTML]{0037D7} +7.2 }        & {\color[HTML]{0037D7} +14.0  }        & {\color[HTML]{0037D7} +77.0  }    &  {\color[HTML]{0037D7} +63.4  }         & {\color[HTML]{0037D7} +97.2 }       \\                                  
                                 \hline\hline
\end{tabular}
  \begin{tablenotes}
    \item[*] *In the Amazon column, B, C, V, T, and S indicate  $Books$, $Clothing$, $Video$, $Toys$, and $Sports$ domain, respectively. In addition, A, C, N, M, and C in the Telco column are the abbreviations for $App$-$Use$, $Call$-$Use$, $Navi$, $Coupon$-$Usage$, and $e$-$comm$ domain, respectively.
    \end{tablenotes}
\end{table}

\vspace{-0.2cm}
\subsection{Discussion of Model Variants (RQ3)} \label{subsection: discussion_model_variants}
To examine the validity of each component, we introduce various variants, each aligned with their specific design purpose (Table \ref{tab:result_4}):
    
\noindent \textbf{(A) \textit{w/o} LC-NTG}: The LC-NTG module is removed in this model variant. Therefore, there is no gradient explicit controller in the training for high NTG domains in the model.
    
\noindent \textbf{(B) \textit{w/o} SC-MIM}: This model variant removes the SC-MIM objective, which means it does not take into account the correlation between single- and cross-domain representations for each domain.

\noindent \textbf{(C) \textit{w/o} ACMoE}: The ACMoE layer is replaced with the vanilla MoE layer without the stop gradient operation.
This variant does not explicitly require experts to be specialized in SDSR and CDSR tasks.
Therefore, it is not possible to compute the pure losses of SDSR and CDSR tasks, and as a result, the exact negative transfer in LC-NTG cannot be obtained.

\begin{table}[]\footnotesize
\centering
\caption{Comparison of MRR for various combinations of components. If the difference between Full SyNCRec and its variants is positive, then it is colored in {\color[HTML]{0037D7} blue}; otherwise {\color[HTML]{CB0000} red}.}
    \vspace{-0.35cm}
    \setlength{\tabcolsep}{1.6pt}
    \renewcommand{\arraystretch}{0.9}
\label{tab:result_4}
\begin{tabular}{l|ccccc|ccccc}
\hline\hline
\multicolumn{1}{c|}{\textbf{Dataset}} & \multicolumn{5}{c|}{\textbf{Amazon}} & \multicolumn{5}{c}{\textbf{Telco}} \\ \hline
\multicolumn{1}{c|}{\textbf{Variants}}       & B     & C     & V     & T     & S    & A     & C     & N    & M    & E    \\ \hline
Full SyNCRec                               &  \textbf{0.248}     &  \textbf{0.258}     & \textbf{0.312}      & \textbf{0.307}      &  \textbf{0.574}    &  \textbf{0.866}     & \textbf{0.651}      & \textbf{0.851}     & \textbf{0.872}     & \textbf{0.443}     \\ 
\hline\hline
(A) \textit{w/o} LC-NTG                &   {\color[HTML]{CB0000} 0.232}    &  {\color[HTML]{0037D7} 0.263}     & {\color[HTML]{CB0000} 0.280}      & {\color[HTML]{CB0000} 0.293}      & {\color[HTML]{CB0000} 0.519}     &  {\color[HTML]{0037D7} 0.891}     &  {\color[HTML]{CB0000} 0.594}     & {\color[HTML]{CB0000} 0.809}     & {\color[HTML]{CB0000} 0.771}     & {\color[HTML]{CB0000} 0.331}     \\
(B) \textit{w/o} SC-MIM                  & {\color[HTML]{CB0000} 0.212}      & {\color[HTML]{CB0000} 0.241}      & {\color[HTML]{CB0000} 0.301}      & {\color[HTML]{0037D7} 0.309}      & {\color[HTML]{CB0000} 0.550}     &  {\color[HTML]{CB0000} 0.865}     &  {\color[HTML]{CB0000} 0.642}     & {\color[HTML]{CB0000} 0.842}     & {\color[HTML]{0037D7} 0.876}     & {\color[HTML]{CB0000} 0.413}     \\
$\llcorner$ CGRec                          & 0.195      & 0.242      & 0.270      & 0.280      & 0.497     &   0.849    & 0.562        & 0.757     & 0.713     & 0.325     \\ 
(C) \textit{w/o} ACMoE                 &   {\color[HTML]{CB0000} 0.220}    & {\color[HTML]{CB0000} 0.237}      &  {\color[HTML]{CB0000} 0.304}     &  {\color[HTML]{CB0000} 0.298}     & {\color[HTML]{CB0000} 0.566}     &   {\color[HTML]{CB0000} 0.836}    & {\color[HTML]{CB0000} 0.647}      & {\color[HTML]{0037D7} 0.857}     & {\color[HTML]{CB0000} 0.855}     & {\color[HTML]{0037D7} 0.445}     \\ 
\hline\hline
\end{tabular}
\end{table}

\noindent \textbf{(1) Each of the three main components of our model contributes to improved performance.}
The prediction performance of our model and its three variants was compared, and it was discovered that removing or replacing any key component resulted in significant performance degradation.
In  particular, the performance gap between full SyNCRec and the (A) $w/o$ LC-NTG variant indicates the advantage of a penalty for the high NTG domain in the training process.
The comparison of full SyNCRec with (B) $w/o$ SC-MIM shows that the SC-MIM module led to performance improvements in most domains, demonstrating the advantage of modeling correlations between single- and cross-domain sequences.
Furthermore, ACMoE effectively decoupled the expert networks to specialize them for SDSR and CDSR tasks. 
This  reflects the accurate negative transfer effect in the model, as the (C) \textit{w/o} ACMoE variant exhibits performance degradation compared to full SyNCRec in most domains in both datasets.
In short, LC-NTG is seen to be more effective when SDSR and CDSR are decoupled via the ACMoE.

\noindent \textbf{(2) Our model, along with its variations, demonstrates enhanced performance when compared to baseline models that share similar design purposes.}
CGRec measures domain-specific NTG using the concept of the Shapley value in terms of the training loss for each domain, and tries loss re-balancing with it.
However, CGRec underperformed relative to the (B) $w/o$ SC-MIM variant (LC-NTG+ACMoE variant) for all domains in both datasets.
In addition, to derive domain-specific NTG, CGRec needs to compute the losses of all combinations of domains, requiring $\mathcal{O}(|\mathcal{D}|!)$ time complexity, whereas our model only computes the losses for two tasks, SDSR and CDSR, requiring only $\mathcal{O}(|\mathcal{D}|)$ time complexity.
This indicates that the proposed loss correction approach is more efficient than the Shapley value based computation in CGRec.

\vspace{-0.3cm}
\subsection{Hyper-parameter Analysis (RQ4)}  \label{subsection:hyperparameter}
\begin{figure}[h]
    \begin{center}
    \includegraphics[width=0.85\linewidth]{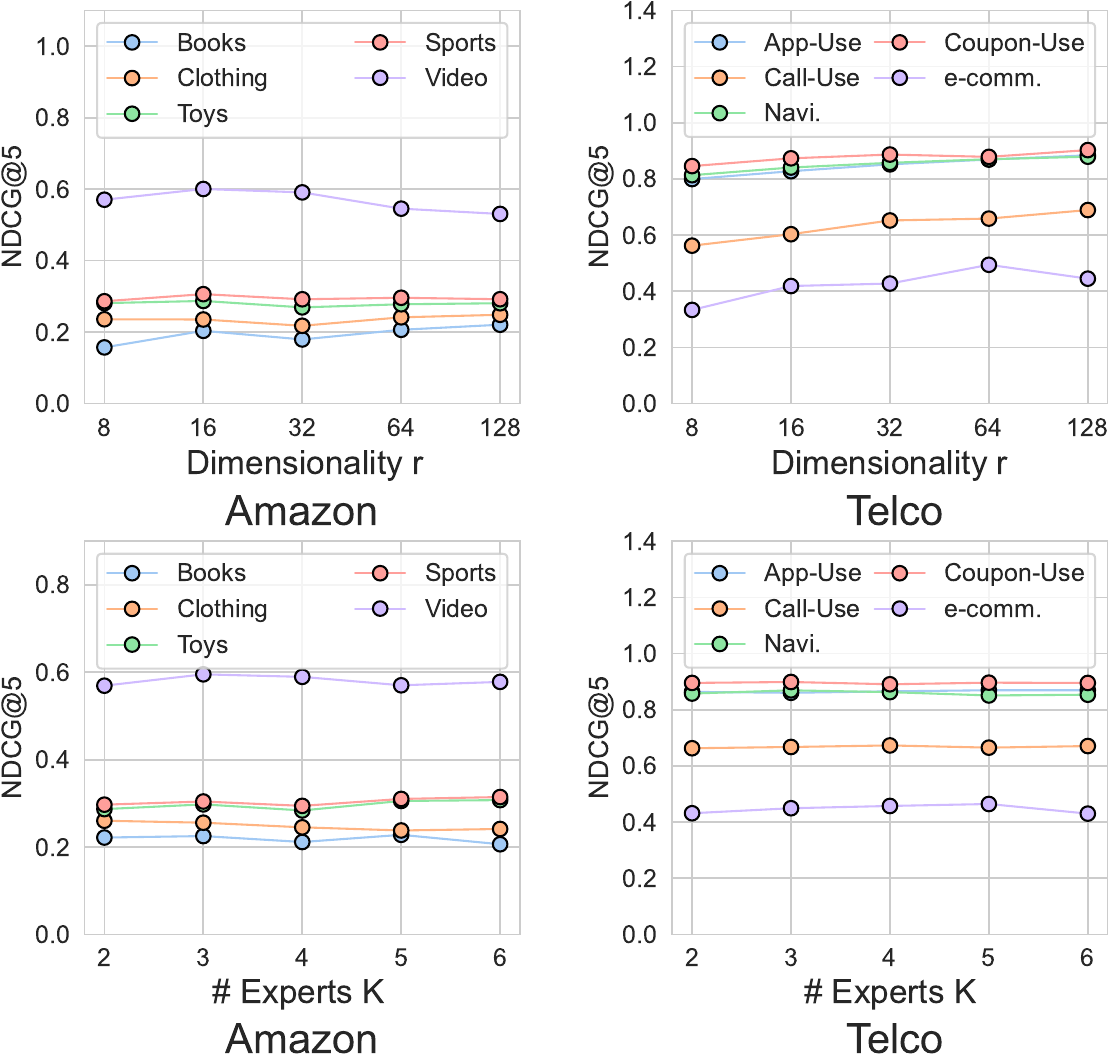}
    \end{center}
    \vspace{-0.3cm}
    \caption{Hyper-parameter study of our model.
    }
    \label{fig:hyper}
    \end{figure}

As shown in Fig. \ref{fig:hyper}, to assess how different hyper-parameter configurations affect our model, we performed experiments on two datasets using various configurations of crucial hyper-parameters, the model dimensionality ($r$) and the number of experts ($K$) in ACMoE. 
The other hyperparameters are set to the same values as those mentioned in Section \ref{subsubsection: hyperparameter_setting}.

\noindent \textbf{(1) Model dimensionality $r$: }
We observed that when $r$ rises from 8 to 16, there is an improvement in the average performance of all domains on the Amazon dataset. 
For the Telco dataset, there is an improvement in the average performance of all domains when $r$ rises from 8 to 64. 
From the results, we confirmed that the threshold for $r$ depends on the characteristics of the datasets. 
The number of interactions of Telco is larger than that of the Amazon dataset, and thus a higher dimensionality is required for the Telco (i.e., 64) than for  the Amazon dataset (i.e., 16) to achieve the best performance.

\noindent \textbf{(2) Number of Sequential Experts $K$: } 
We experimented with a range of experts in ACMoE, from 2 to 6, and the results are presented in Fig. \ref{fig:hyper}. 
Under all tested settings, our approach consistently surpassed the best baseline model (second best for each domain), in most domains of the two datasets.
This result underscores the robustness of our approach with respect to the configuration of the mixture of experts. 

We also verified the effects of the number of decoupled experts ($j$) for CDSR in ACMoE (Section \ref{subsubsection:mose}).
In both datasets, when three ($j$) out of five experts ($K$) are trained by the CDSR task (i.e., the stop gradient is applied to two experts for the CDSR task), the highest performance is achieved for all domains. 
This indicates that assigning more experts to the challenging CDSR task improves model performance.
Due to limited space, a detailed analysis of $j$ cannot be provided in this section. 

    \vspace{-0.2cm}
    
\section{Online A/B Test (RQ 5)}
We deployed our recommendation model in the personal assistant service of our company. 
Online experiments were conducted within an A/B testing framework from October 2023 to January 2024. The serving model used during this period was a variant of our model, adapted to suit business characteristics.
Our model was trained on sequential logs of approximately 12 million users from nine domains, including the personal assistant service domain.
It recommends the \textit{Top}@1 next item within the personal assistant service domain to the customer.
Our primary metric for online evaluation is the Click-Through Rate (CTR), defined as the ratio of clicks to impressions. 
Results from the online A/B test revealed that our model achieved 17.3\% and 25.6\% increases in online CTR compared to the SASRec-based \cite{kang2018self} model and rule-based model, respectively, affirming the practical effectiveness of our model.

\section{Conclusion}
We introduce a CDSR framework that tackles negative transfer by dynamically applying a weight to the prediction loss, which is determined based on the estimated evaluation of negative transfer.
Additionally, we developed an auxiliary loss to enhance the exchange of valuable information between SDSR and CDSR tasks on a per-domain basis.
Our model outperformed existing models in extensive tests on two real-world datasets with ten domains. Its implementation in our personal assistant app's recommendation system yielded a significant increase in click-through rate, demonstrating substantial business value.

\clearpage
\bibliographystyle{ACM-Reference-Format}
\balance
\bibliography{my_ref}


\begin{thebibliography}{38}


\ifx \showCODEN    \undefined \def \showCODEN     #1{\unskip}     \fi
\ifx \showDOI      \undefined \def \showDOI       #1{#1}\fi
\ifx \showISBNx    \undefined \def \showISBNx     #1{\unskip}     \fi
\ifx \showISBNxiii \undefined \def \showISBNxiii  #1{\unskip}     \fi
\ifx \showISSN     \undefined \def \showISSN      #1{\unskip}     \fi
\ifx \showLCCN     \undefined \def \showLCCN      #1{\unskip}     \fi
\ifx \shownote     \undefined \def \shownote      #1{#1}          \fi
\ifx \showarticletitle \undefined \def \showarticletitle #1{#1}   \fi
\ifx \showURL      \undefined \def \showURL       {\relax}        \fi
\providecommand\bibfield[2]{#2}
\providecommand\bibinfo[2]{#2}
\providecommand\natexlab[1]{#1}
\providecommand\showeprint[2][]{arXiv:#2}

\bibitem[Ba et~al\mbox{.}(2016)]%
        {ba2016layer}
\bibfield{author}{\bibinfo{person}{Jimmy~Lei Ba}, \bibinfo{person}{Jamie~Ryan Kiros}, {and} \bibinfo{person}{Geoffrey~E Hinton}.} \bibinfo{year}{2016}\natexlab{}.
\newblock \showarticletitle{Layer normalization}.
\newblock \bibinfo{journal}{\emph{arXiv preprint arXiv:1607.06450}} (\bibinfo{year}{2016}).
\newblock


\bibitem[Berg et~al\mbox{.}(2017)]%
        {berg2017graph}
\bibfield{author}{\bibinfo{person}{Rianne van~den Berg}, \bibinfo{person}{Thomas~N Kipf}, {and} \bibinfo{person}{Max Welling}.} \bibinfo{year}{2017}\natexlab{}.
\newblock \showarticletitle{Graph convolutional matrix completion}.
\newblock \bibinfo{journal}{\emph{arXiv preprint arXiv:1706.02263}} (\bibinfo{year}{2017}).
\newblock


\bibitem[Cao et~al\mbox{.}(2022)]%
        {cao2022contrastive}
\bibfield{author}{\bibinfo{person}{Jiangxia Cao}, \bibinfo{person}{Xin Cong}, \bibinfo{person}{Jiawei Sheng}, \bibinfo{person}{Tingwen Liu}, {and} \bibinfo{person}{Bin Wang}.} \bibinfo{year}{2022}\natexlab{}.
\newblock \showarticletitle{Contrastive Cross-Domain Sequential Recommendation}. In \bibinfo{booktitle}{\emph{Proceedings of the 31st ACM International Conference on Information \& Knowledge Management}}. \bibinfo{pages}{138--147}.
\newblock


\bibitem[Fan et~al\mbox{.}(2021)]%
        {fan2021lighter}
\bibfield{author}{\bibinfo{person}{Xinyan Fan}, \bibinfo{person}{Zheng Liu}, \bibinfo{person}{Jianxun Lian}, \bibinfo{person}{Wayne~Xin Zhao}, \bibinfo{person}{Xing Xie}, {and} \bibinfo{person}{Ji-Rong Wen}.} \bibinfo{year}{2021}\natexlab{}.
\newblock \showarticletitle{Lighter and better: low-rank decomposed self-attention networks for next-item recommendation}. In \bibinfo{booktitle}{\emph{Proceedings of the 44th international ACM SIGIR conference on research and development in information retrieval}}. \bibinfo{pages}{1733--1737}.
\newblock


\bibitem[Gao et~al\mbox{.}(2013)]%
        {gao2013cross}
\bibfield{author}{\bibinfo{person}{Sheng Gao}, \bibinfo{person}{Hao Luo}, \bibinfo{person}{Da Chen}, \bibinfo{person}{Shantao Li}, \bibinfo{person}{Patrick Gallinari}, {and} \bibinfo{person}{Jun Guo}.} \bibinfo{year}{2013}\natexlab{}.
\newblock \showarticletitle{Cross-domain recommendation via cluster-level latent factor model}. In \bibinfo{booktitle}{\emph{Machine Learning and Knowledge Discovery in Databases: European Conference, ECML PKDD 2013, Prague, Czech Republic, September 23-27, 2013, Proceedings, Part II 13}}. Springer, \bibinfo{pages}{161--176}.
\newblock


\bibitem[He et~al\mbox{.}(2017)]%
        {he2017translation}
\bibfield{author}{\bibinfo{person}{Ruining He}, \bibinfo{person}{Wang-Cheng Kang}, {and} \bibinfo{person}{Julian McAuley}.} \bibinfo{year}{2017}\natexlab{}.
\newblock \showarticletitle{Translation-based recommendation}. In \bibinfo{booktitle}{\emph{Proceedings of the eleventh ACM conference on recommender systems}}. \bibinfo{pages}{161--169}.
\newblock


\bibitem[Hendrycks and Gimpel(2016)]%
        {hendrycks2016gaussian}
\bibfield{author}{\bibinfo{person}{Dan Hendrycks} {and} \bibinfo{person}{Kevin Gimpel}.} \bibinfo{year}{2016}\natexlab{}.
\newblock \showarticletitle{Gaussian error linear units (gelus)}.
\newblock \bibinfo{journal}{\emph{arXiv preprint arXiv:1606.08415}} (\bibinfo{year}{2016}).
\newblock


\bibitem[Hidasi et~al\mbox{.}(2015)]%
        {hidasi2015session}
\bibfield{author}{\bibinfo{person}{Bal{\'a}zs Hidasi}, \bibinfo{person}{Alexandros Karatzoglou}, \bibinfo{person}{Linas Baltrunas}, {and} \bibinfo{person}{Domonkos Tikk}.} \bibinfo{year}{2015}\natexlab{}.
\newblock \showarticletitle{Session-based recommendations with recurrent neural networks}.
\newblock \bibinfo{journal}{\emph{arXiv preprint arXiv:1511.06939}} (\bibinfo{year}{2015}).
\newblock


\bibitem[Kang and McAuley(2018)]%
        {kang2018self}
\bibfield{author}{\bibinfo{person}{Wang-Cheng Kang} {and} \bibinfo{person}{Julian McAuley}.} \bibinfo{year}{2018}\natexlab{}.
\newblock \showarticletitle{Self-attentive sequential recommendation}. In \bibinfo{booktitle}{\emph{2018 IEEE international conference on data mining (ICDM)}}. IEEE, \bibinfo{pages}{197--206}.
\newblock


\bibitem[Kong et~al\mbox{.}(2019)]%
        {kong2019mutual}
\bibfield{author}{\bibinfo{person}{Lingpeng Kong}, \bibinfo{person}{Cyprien de~Masson d'Autume}, \bibinfo{person}{Wang Ling}, \bibinfo{person}{Lei Yu}, \bibinfo{person}{Zihang Dai}, {and} \bibinfo{person}{Dani Yogatama}.} \bibinfo{year}{2019}\natexlab{}.
\newblock \showarticletitle{A mutual information maximization perspective of language representation learning}.
\newblock \bibinfo{journal}{\emph{arXiv preprint arXiv:1910.08350}} (\bibinfo{year}{2019}).
\newblock


\bibitem[Li et~al\mbox{.}(2023)]%
        {li2023one}
\bibfield{author}{\bibinfo{person}{Chenglin Li}, \bibinfo{person}{Yuanzhen Xie}, \bibinfo{person}{Chenyun Yu}, \bibinfo{person}{Bo Hu}, \bibinfo{person}{Zang Li}, \bibinfo{person}{Guoqiang Shu}, \bibinfo{person}{Xiaohu Qie}, {and} \bibinfo{person}{Di Niu}.} \bibinfo{year}{2023}\natexlab{}.
\newblock \showarticletitle{One for All, All for One: Learning and Transferring User Embeddings for Cross-Domain Recommendation}. In \bibinfo{booktitle}{\emph{Proceedings of the Sixteenth ACM International Conference on Web Search and Data Mining}}. \bibinfo{pages}{366--374}.
\newblock


\bibitem[Li et~al\mbox{.}(2017)]%
        {li2017neural}
\bibfield{author}{\bibinfo{person}{Jing Li}, \bibinfo{person}{Pengjie Ren}, \bibinfo{person}{Zhumin Chen}, \bibinfo{person}{Zhaochun Ren}, \bibinfo{person}{Tao Lian}, {and} \bibinfo{person}{Jun Ma}.} \bibinfo{year}{2017}\natexlab{}.
\newblock \showarticletitle{Neural attentive session-based recommendation}. In \bibinfo{booktitle}{\emph{Proceedings of the 2017 ACM on Conference on Information and Knowledge Management}}. \bibinfo{pages}{1419--1428}.
\newblock


\bibitem[Lin et~al\mbox{.}(2023)]%
        {lin2023mixed}
\bibfield{author}{\bibinfo{person}{Guanyu Lin}, \bibinfo{person}{Chen Gao}, \bibinfo{person}{Yu Zheng}, \bibinfo{person}{Jianxin Chang}, \bibinfo{person}{Yanan Niu}, \bibinfo{person}{Yang Song}, \bibinfo{person}{Kun Gai}, \bibinfo{person}{Zhiheng Li}, \bibinfo{person}{Depeng Jin}, \bibinfo{person}{Yong Li}, {et~al\mbox{.}}} \bibinfo{year}{2023}\natexlab{}.
\newblock \showarticletitle{Mixed Attention Network for Cross-domain Sequential Recommendation}.
\newblock \bibinfo{journal}{\emph{arXiv preprint arXiv:2311.08272}} (\bibinfo{year}{2023}).
\newblock


\bibitem[Liu et~al\mbox{.}(2020)]%
        {liu2020cross}
\bibfield{author}{\bibinfo{person}{Meng Liu}, \bibinfo{person}{Jianjun Li}, \bibinfo{person}{Guohui Li}, {and} \bibinfo{person}{Peng Pan}.} \bibinfo{year}{2020}\natexlab{}.
\newblock \showarticletitle{Cross domain recommendation via bi-directional transfer graph collaborative filtering networks}. In \bibinfo{booktitle}{\emph{Proceedings of the 29th ACM international conference on information \& knowledge management}}. \bibinfo{pages}{885--894}.
\newblock


\bibitem[Liu et~al\mbox{.}(2018)]%
        {liu2018stamp}
\bibfield{author}{\bibinfo{person}{Qiao Liu}, \bibinfo{person}{Yifu Zeng}, \bibinfo{person}{Refuoe Mokhosi}, {and} \bibinfo{person}{Haibin Zhang}.} \bibinfo{year}{2018}\natexlab{}.
\newblock \showarticletitle{STAMP: short-term attention/memory priority model for session-based recommendation}. In \bibinfo{booktitle}{\emph{Proceedings of the 24th ACM SIGKDD international conference on knowledge discovery \& data mining}}. \bibinfo{pages}{1831--1839}.
\newblock


\bibitem[Ma et~al\mbox{.}(2022)]%
        {ma2022mixed}
\bibfield{author}{\bibinfo{person}{Muyang Ma}, \bibinfo{person}{Pengjie Ren}, \bibinfo{person}{Zhumin Chen}, \bibinfo{person}{Zhaochun Ren}, \bibinfo{person}{Lifan Zhao}, \bibinfo{person}{Peiyu Liu}, \bibinfo{person}{Jun Ma}, {and} \bibinfo{person}{Maarten de Rijke}.} \bibinfo{year}{2022}\natexlab{}.
\newblock \showarticletitle{Mixed information flow for cross-domain sequential recommendations}.
\newblock \bibinfo{journal}{\emph{ACM Transactions on Knowledge Discovery from Data (TKDD)}} \bibinfo{volume}{16}, \bibinfo{number}{4} (\bibinfo{year}{2022}), \bibinfo{pages}{1--32}.
\newblock


\bibitem[Ma et~al\mbox{.}(2019)]%
        {ma2019pi}
\bibfield{author}{\bibinfo{person}{Muyang Ma}, \bibinfo{person}{Pengjie Ren}, \bibinfo{person}{Yujie Lin}, \bibinfo{person}{Zhumin Chen}, \bibinfo{person}{Jun Ma}, {and} \bibinfo{person}{Maarten~de Rijke}.} \bibinfo{year}{2019}\natexlab{}.
\newblock \showarticletitle{$\pi$-net: A parallel information-sharing network for shared-account cross-domain sequential recommendations}. In \bibinfo{booktitle}{\emph{Proceedings of the 42nd international ACM SIGIR conference on research and development in information retrieval}}. \bibinfo{pages}{685--694}.
\newblock


\bibitem[McAuley et~al\mbox{.}(2015)]%
        {mcauley2015image}
\bibfield{author}{\bibinfo{person}{Julian McAuley}, \bibinfo{person}{Christopher Targett}, \bibinfo{person}{Qinfeng Shi}, {and} \bibinfo{person}{Anton Van Den~Hengel}.} \bibinfo{year}{2015}\natexlab{}.
\newblock \showarticletitle{Image-based recommendations on styles and substitutes}. In \bibinfo{booktitle}{\emph{Proceedings of the 38th international ACM SIGIR conference on research and development in information retrieval}}. \bibinfo{pages}{43--52}.
\newblock


\bibitem[Oord et~al\mbox{.}(2018)]%
        {oord2018representation}
\bibfield{author}{\bibinfo{person}{Aaron van~den Oord}, \bibinfo{person}{Yazhe Li}, {and} \bibinfo{person}{Oriol Vinyals}.} \bibinfo{year}{2018}\natexlab{}.
\newblock \showarticletitle{Representation learning with contrastive predictive coding}.
\newblock \bibinfo{journal}{\emph{arXiv preprint arXiv:1807.03748}} (\bibinfo{year}{2018}).
\newblock


\bibitem[Park et~al\mbox{.}(2023)]%
        {park2023cracking}
\bibfield{author}{\bibinfo{person}{Chung Park}, \bibinfo{person}{Taesan Kim}, \bibinfo{person}{Taekyoon Choi}, \bibinfo{person}{Junui Hong}, \bibinfo{person}{Yelim Yu}, \bibinfo{person}{Mincheol Cho}, \bibinfo{person}{Kyunam Lee}, \bibinfo{person}{Sungil Ryu}, \bibinfo{person}{Hyungjun Yoon}, \bibinfo{person}{Minsung Choi}, {et~al\mbox{.}}} \bibinfo{year}{2023}\natexlab{}.
\newblock \showarticletitle{Cracking the Code of Negative Transfer: A Cooperative Game Theoretic Approach for Cross-Domain Sequential Recommendation}. In \bibinfo{booktitle}{\emph{Proceedings of the 32nd ACM International Conference on Information and Knowledge Management}}. \bibinfo{pages}{2024--2033}.
\newblock


\bibitem[Qin et~al\mbox{.}(2020)]%
        {qin2020multitask}
\bibfield{author}{\bibinfo{person}{Zhen Qin}, \bibinfo{person}{Yicheng Cheng}, \bibinfo{person}{Zhe Zhao}, \bibinfo{person}{Zhe Chen}, \bibinfo{person}{Donald Metzler}, {and} \bibinfo{person}{Jingzheng Qin}.} \bibinfo{year}{2020}\natexlab{}.
\newblock \showarticletitle{Multitask mixture of sequential experts for user activity streams}. In \bibinfo{booktitle}{\emph{Proceedings of the 26th ACM SIGKDD International Conference on Knowledge Discovery \& Data Mining}}. \bibinfo{pages}{3083--3091}.
\newblock


\bibitem[Rashed et~al\mbox{.}(2022)]%
        {rashed2022context}
\bibfield{author}{\bibinfo{person}{Ahmed Rashed}, \bibinfo{person}{Shereen Elsayed}, {and} \bibinfo{person}{Lars Schmidt-Thieme}.} \bibinfo{year}{2022}\natexlab{}.
\newblock \showarticletitle{Context and attribute-aware sequential recommendation via cross-attention}. In \bibinfo{booktitle}{\emph{Proceedings of the 16th ACM Conference on Recommender Systems}}. \bibinfo{pages}{71--80}.
\newblock


\bibitem[Rendle et~al\mbox{.}(2012)]%
        {rendle2012bpr}
\bibfield{author}{\bibinfo{person}{Steffen Rendle}, \bibinfo{person}{Christoph Freudenthaler}, \bibinfo{person}{Zeno Gantner}, {and} \bibinfo{person}{Lars Schmidt-Thieme}.} \bibinfo{year}{2012}\natexlab{}.
\newblock \showarticletitle{BPR: Bayesian personalized ranking from implicit feedback}.
\newblock \bibinfo{journal}{\emph{arXiv preprint arXiv:1205.2618}} (\bibinfo{year}{2012}).
\newblock


\bibitem[Singh and Gordon(2008)]%
        {singh2008relational}
\bibfield{author}{\bibinfo{person}{Ajit~P Singh} {and} \bibinfo{person}{Geoffrey~J Gordon}.} \bibinfo{year}{2008}\natexlab{}.
\newblock \showarticletitle{Relational learning via collective matrix factorization}. In \bibinfo{booktitle}{\emph{Proceedings of the 14th ACM SIGKDD international conference on Knowledge discovery and data mining}}. \bibinfo{pages}{650--658}.
\newblock


\bibitem[Sun et~al\mbox{.}(2019)]%
        {sun2019bert4rec}
\bibfield{author}{\bibinfo{person}{Fei Sun}, \bibinfo{person}{Jun Liu}, \bibinfo{person}{Jian Wu}, \bibinfo{person}{Changhua Pei}, \bibinfo{person}{Xiao Lin}, \bibinfo{person}{Wenwu Ou}, {and} \bibinfo{person}{Peng Jiang}.} \bibinfo{year}{2019}\natexlab{}.
\newblock \showarticletitle{BERT4Rec: Sequential recommendation with bidirectional encoder representations from transformer}. In \bibinfo{booktitle}{\emph{Proceedings of the 28th ACM international conference on information and knowledge management}}. \bibinfo{pages}{1441--1450}.
\newblock


\bibitem[Szegedy et~al\mbox{.}(2016)]%
        {szegedy2016rethinking}
\bibfield{author}{\bibinfo{person}{Christian Szegedy}, \bibinfo{person}{Vincent Vanhoucke}, \bibinfo{person}{Sergey Ioffe}, \bibinfo{person}{Jon Shlens}, {and} \bibinfo{person}{Zbigniew Wojna}.} \bibinfo{year}{2016}\natexlab{}.
\newblock \showarticletitle{Rethinking the inception architecture for computer vision}. In \bibinfo{booktitle}{\emph{Proceedings of the IEEE conference on computer vision and pattern recognition}}. \bibinfo{pages}{2818--2826}.
\newblock


\bibitem[Tan et~al\mbox{.}(2021)]%
        {tan2021sparse}
\bibfield{author}{\bibinfo{person}{Qiaoyu Tan}, \bibinfo{person}{Jianwei Zhang}, \bibinfo{person}{Jiangchao Yao}, \bibinfo{person}{Ninghao Liu}, \bibinfo{person}{Jingren Zhou}, \bibinfo{person}{Hongxia Yang}, {and} \bibinfo{person}{Xia Hu}.} \bibinfo{year}{2021}\natexlab{}.
\newblock \showarticletitle{Sparse-interest network for sequential recommendation}. In \bibinfo{booktitle}{\emph{Proceedings of the 14th ACM international conference on web search and data mining}}. \bibinfo{pages}{598--606}.
\newblock


\bibitem[Vaswani et~al\mbox{.}(2017)]%
        {vaswani2017attention}
\bibfield{author}{\bibinfo{person}{Ashish Vaswani}, \bibinfo{person}{Noam Shazeer}, \bibinfo{person}{Niki Parmar}, \bibinfo{person}{Jakob Uszkoreit}, \bibinfo{person}{Llion Jones}, \bibinfo{person}{Aidan~N Gomez}, \bibinfo{person}{{\L}ukasz Kaiser}, {and} \bibinfo{person}{Illia Polosukhin}.} \bibinfo{year}{2017}\natexlab{}.
\newblock \showarticletitle{Attention is all you need}. In \bibinfo{booktitle}{\emph{Advances in neural information processing systems}}. \bibinfo{pages}{5998--6008}.
\newblock


\bibitem[Wang et~al\mbox{.}(2019)]%
        {wang2019characterizing}
\bibfield{author}{\bibinfo{person}{Zirui Wang}, \bibinfo{person}{Zihang Dai}, \bibinfo{person}{Barnab{\'a}s P{\'o}czos}, {and} \bibinfo{person}{Jaime Carbonell}.} \bibinfo{year}{2019}\natexlab{}.
\newblock \showarticletitle{Characterizing and avoiding negative transfer}. In \bibinfo{booktitle}{\emph{Proceedings of the IEEE/CVF conference on computer vision and pattern recognition}}. \bibinfo{pages}{11293--11302}.
\newblock


\bibitem[Yan et~al\mbox{.}(2019)]%
        {yan2019deepapf}
\bibfield{author}{\bibinfo{person}{Huan Yan}, \bibinfo{person}{Xiangning Chen}, \bibinfo{person}{Chen Gao}, \bibinfo{person}{Yong Li}, {and} \bibinfo{person}{Depeng Jin}.} \bibinfo{year}{2019}\natexlab{}.
\newblock \showarticletitle{Deepapf: Deep attentive probabilistic factorization for multi-site video recommendation}.
\newblock \bibinfo{journal}{\emph{TC}} \bibinfo{volume}{2}, \bibinfo{number}{130} (\bibinfo{year}{2019}), \bibinfo{pages}{17--883}.
\newblock


\bibitem[Yuan et~al\mbox{.}(2019)]%
        {yuan2019simple}
\bibfield{author}{\bibinfo{person}{Fajie Yuan}, \bibinfo{person}{Alexandros Karatzoglou}, \bibinfo{person}{Ioannis Arapakis}, \bibinfo{person}{Joemon~M Jose}, {and} \bibinfo{person}{Xiangnan He}.} \bibinfo{year}{2019}\natexlab{}.
\newblock \showarticletitle{A simple convolutional generative network for next item recommendation}. In \bibinfo{booktitle}{\emph{Proceedings of the twelfth ACM international conference on web search and data mining}}. \bibinfo{pages}{582--590}.
\newblock


\bibitem[Zhang et~al\mbox{.}(2019)]%
        {zhang2019feature}
\bibfield{author}{\bibinfo{person}{Tingting Zhang}, \bibinfo{person}{Pengpeng Zhao}, \bibinfo{person}{Yanchi Liu}, \bibinfo{person}{Victor~S Sheng}, \bibinfo{person}{Jiajie Xu}, \bibinfo{person}{Deqing Wang}, \bibinfo{person}{Guanfeng Liu}, \bibinfo{person}{Xiaofang Zhou}, {et~al\mbox{.}}} \bibinfo{year}{2019}\natexlab{}.
\newblock \showarticletitle{Feature-level Deeper Self-Attention Network for Sequential Recommendation.}. In \bibinfo{booktitle}{\emph{IJCAI}}. \bibinfo{pages}{4320--4326}.
\newblock


\bibitem[Zhang et~al\mbox{.}(2023)]%
        {zhang2023collaborative}
\bibfield{author}{\bibinfo{person}{Wei Zhang}, \bibinfo{person}{Pengye Zhang}, \bibinfo{person}{Bo Zhang}, \bibinfo{person}{Xingxing Wang}, {and} \bibinfo{person}{Dong Wang}.} \bibinfo{year}{2023}\natexlab{}.
\newblock \showarticletitle{A Collaborative Transfer Learning Framework for Cross-domain Recommendation}. In \bibinfo{booktitle}{\emph{Proceedings of the 29th ACM SIGKDD Conference on Knowledge Discovery and Data Mining}}. \bibinfo{pages}{5576--5585}.
\newblock


\bibitem[Zhao et~al\mbox{.}(2022)]%
        {recbole[2.0]}
\bibfield{author}{\bibinfo{person}{Wayne~Xin Zhao}, \bibinfo{person}{Yupeng Hou}, \bibinfo{person}{Xingyu Pan}, \bibinfo{person}{Chen Yang}, \bibinfo{person}{Zeyu Zhang}, \bibinfo{person}{Zihan Lin}, \bibinfo{person}{Jingsen Zhang}, \bibinfo{person}{Shuqing Bian}, \bibinfo{person}{Jiakai Tang}, \bibinfo{person}{Wenqi Sun}, {et~al\mbox{.}}} \bibinfo{year}{2022}\natexlab{}.
\newblock \showarticletitle{RecBole 2.0: Towards a More Up-to-Date Recommendation Library}. In \bibinfo{booktitle}{\emph{Proceedings of the 31st ACM International Conference on Information \& Knowledge Management}}. \bibinfo{pages}{4722--4726}.
\newblock


\bibitem[Zhao et~al\mbox{.}(2021)]%
        {zhao2021recbole}
\bibfield{author}{\bibinfo{person}{Wayne~Xin Zhao}, \bibinfo{person}{Shanlei Mu}, \bibinfo{person}{Yupeng Hou}, \bibinfo{person}{Zihan Lin}, \bibinfo{person}{Kaiyuan Li}, \bibinfo{person}{Yushuo Chen}, \bibinfo{person}{Yujie Lu}, \bibinfo{person}{Hui Wang}, \bibinfo{person}{Changxin Tian}, \bibinfo{person}{Xingyu Pan}, \bibinfo{person}{Yingqian Min}, \bibinfo{person}{Zhichao Feng}, \bibinfo{person}{Xinyan Fan}, \bibinfo{person}{Xu Chen}, \bibinfo{person}{Pengfei Wang}, \bibinfo{person}{Wendi Ji}, \bibinfo{person}{Yaliang Li}, \bibinfo{person}{Xiaoling Wang}, {and} \bibinfo{person}{Ji-Rong Wen}.} \bibinfo{year}{2021}\natexlab{}.
\newblock \showarticletitle{Recbole: Towards a unified, comprehensive and efficient framework for recommendation algorithms}. In \bibinfo{booktitle}{\emph{{CIKM}}}.
\newblock


\bibitem[Zhou et~al\mbox{.}(2020)]%
        {zhou2020s3}
\bibfield{author}{\bibinfo{person}{Kun Zhou}, \bibinfo{person}{Hui Wang}, \bibinfo{person}{Wayne~Xin Zhao}, \bibinfo{person}{Yutao Zhu}, \bibinfo{person}{Sirui Wang}, \bibinfo{person}{Fuzheng Zhang}, \bibinfo{person}{Zhongyuan Wang}, {and} \bibinfo{person}{Ji-Rong Wen}.} \bibinfo{year}{2020}\natexlab{}.
\newblock \showarticletitle{S3-rec: Self-supervised learning for sequential recommendation with mutual information maximization}. In \bibinfo{booktitle}{\emph{Proceedings of the 29th ACM international conference on information \& knowledge management}}. \bibinfo{pages}{1893--1902}.
\newblock


\bibitem[Zhu et~al\mbox{.}(2019)]%
        {zhu2019dtcdr}
\bibfield{author}{\bibinfo{person}{Feng Zhu}, \bibinfo{person}{Chaochao Chen}, \bibinfo{person}{Yan Wang}, \bibinfo{person}{Guanfeng Liu}, {and} \bibinfo{person}{Xiaolin Zheng}.} \bibinfo{year}{2019}\natexlab{}.
\newblock \showarticletitle{Dtcdr: A framework for dual-target cross-domain recommendation}. In \bibinfo{booktitle}{\emph{Proceedings of the 28th ACM International Conference on Information and Knowledge Management}}. \bibinfo{pages}{1533--1542}.
\newblock


\bibitem[Zhu et~al\mbox{.}(2021)]%
        {zhu2021cross}
\bibfield{author}{\bibinfo{person}{Feng Zhu}, \bibinfo{person}{Yan Wang}, \bibinfo{person}{Chaochao Chen}, \bibinfo{person}{Jun Zhou}, \bibinfo{person}{Longfei Li}, {and} \bibinfo{person}{Guanfeng Liu}.} \bibinfo{year}{2021}\natexlab{}.
\newblock \showarticletitle{Cross-domain recommendation: challenges, progress, and prospects}.
\newblock \bibinfo{journal}{\emph{arXiv preprint arXiv:2103.01696}} (\bibinfo{year}{2021}).
\newblock


\end{thebibliography}

\appendix

\end{document}